\documentclass[fleqn,usenatbib]{mnras}

\usepackage{newtxtext,newtxmath}

\usepackage[T1]{fontenc}

\DeclareRobustCommand{\VAN}[3]{#2}
\let\VANthebibliography\thebibliography
\def\thebibliography{\DeclareRobustCommand{\VAN}[3]{##3}\VANthebibliography}

\usepackage{graphicx}	
\usepackage{amsmath}	

\usepackage{scalerel,tikz}

\usepackage{float}

\usetikzlibrary{svg.path}
\definecolor{orcidlogocol}{HTML}{A6CE39}
\tikzset{orcidlogo/.pic={
 \fill[orcidlogocol] svg{M256,128c0,70.7-57.3,128-128,128C57.3,256,0,198.7,0,128C0,57.3,57.3,0,128,0C198.7,0,256,57.3,256,128z};
 \fill[white] svg{M86.3,186.2H70.9V79.1h15.4v48.4V186.2z}
 svg{M108.9,79.1h41.6c39.6,0,57,28.3,57,53.6c0,27.5-21.5,53.6-56.8,53.6h-41.8V79.1z M124.3,172.4h24.5c34.9,0,42.9-26.5,42.9-39.7c0-21.5-13.7-39.7-43.7-39.7h-23.7V172.4z}
 svg{M88.7,56.8c0,5.5-4.5,10.1-10.1,10.1c-5.6,0-10.1-4.6-10.1-10.1c0-5.6,4.5-10.1,10.1-10.1C84.2,46.7,88.7,51.3,88.7,56.8z};
}}
\newcommand\orcidicon[1]{\href{https://orcid.org/#1}{\mbox{\scalerel*{
\begin{tikzpicture}[yscale=-1,transform shape]
\pic{orcidlogo};
\end{tikzpicture}
}{|}}}}

\title[Shock-driven turbulence]{The driving mode of shock-driven turbulence}

\author[S.~Dhawalikar et al.]{
Saee Dhawalikar$^{\orcidicon{0000-0002-8454-7622}\,1,2}$\thanks{E-mail: saee.dhawalikar@iucaa.in (SD)},
Christoph Federrath$^{\orcidicon{0000-0002-0706-2306}\,1,3}$\thanks{E-mail: christoph.federrath@anu.edu.au (CF)},
Seth Davidovits$^{\orcidicon{0000-0002-4808-7286}\,4}$\thanks{E-mail: davidovits1@llnl.gov (SD)},
Romain Teyssier$^{\orcidicon{0000-0001-7689-0933}\,5}$\thanks{E-mail: teyssier@princeton.edu (RT)},
\newauthor\hspace{0.02cm}
Sabrina R.~Nagel$^{\orcidicon{0000-0002-7768-6819}\,4}$\thanks{E-mail: nagel7@llnl.gov (SRN)},
Bruce A.~Remington$^{\orcidicon{0000-0002-9369-7040}\,4}$\thanks{E-mail: remington2@llnl.gov (BAR)}, and
David C.~Collins$^{\orcidicon{0000-0001-6661-2243}\,6}$\thanks{E-mail: dccollins@fsu.edu (DCC)}
\vspace{0.1cm}
\\
$^{1}$Research School of Astronomy and Astrophysics, Australian National University, Canberra, ACT~2611, Australia\\
$^{2}$Inter-University Centre for Astronomy and Astrophysics (IUCAA), Pune 411007, India\\
$^{3}$Australian Research Council Centre of Excellence in All Sky Astrophysics (ASTRO3D), Canberra, ACT~2611, Australia\\
$^{4}$Lawrence Livermore National Laboratory, Livermore, California 94550, USA\\
$^{5}$Department of Astrophysical Sciences, Princeton University, 4~Ivy Lane, 08544, Princeton, NJ, USA\\
$^{6}$Department of Physics, Florida State University, Tallahassee, FL~32306-4350, USA
}

\date{Accepted XXX. Received YYY; in original form ZZZ}

\pubyear{2015}

\begin{document}
\label{firstpage}
\pagerange{\pageref{firstpage}--\pageref{lastpage}}
\maketitle

\begin{abstract}
Turbulence in the interstellar medium (ISM) is crucial in the process of star formation. Shocks produced by supernova explosions, jets, radiation from massive stars, or galactic spiral-arm dynamics are amongst the most common drivers of turbulence in the ISM. However, it is not fully understood how shocks drive turbulence, in particular whether shock driving is a more solenoidal (rotational, divergence-free) or a more compressive (potential, curl-free) mode of driving turbulence. The mode of turbulence driving has profound consequences for star formation, with compressive driving producing three times larger density dispersion, and an order of magnitude higher star formation rate than solenoidal driving. Here, we use hydrodynamical simulations of a shock inducing turbulent motions in a structured, multi-phase medium. This is done in the context of a laser-induced shock, propagating into a foam material, in preparation for an experiment to be performed at the National Ignition Facility (NIF). Specifically, we analyse the density and velocity distributions in the shocked turbulent medium, and measure the turbulence driving parameter $b=(\sigma_{\rho/\langle\rho\rangle}^{2\Gamma}-1)^{1/2}(1-\sigma_{\rho/\langle\rho\rangle}^{-2})^{-1/2}\mathcal{M}^{-1}\Gamma^{-1/2}$, with the density dispersion $\sigma_{\rho/\langle\rho\rangle}$, the turbulent Mach number $\mathcal{M}$, and the polytropic exponent $\Gamma$. Purely solenoidal and purely compressive driving correspond to $b\sim1/3$ and $b\sim1$, respectively. Using simulations in which a shock is driven into a multi-phase medium with structures of different sizes and $\Gamma<1$, we find $b\sim1$ for all cases, showing that shock-driven turbulence is consistent with strongly compressive driving.
\end{abstract}

\begin{keywords}
hydrodynamics -- instabilities -- shock waves -- turbulence
\end{keywords}



\section{Introduction}
Turbulence in the interstellar medium plays a key role in the process of star formation \citep{MacLowKlessen2004,ElmegreenScalo2004,McKeeOstriker2007,HennebelleFalgarone2012,PadoanEtAl2014}. The statistics of turbulence can be used to determine the star formation rate (SRF) \citep{KrumholzMcKee2005,PadoanNordlund2011,HennebelleChabrier2011,FederrathKlessen2012}, as well as the initial mass function (IMF) of stars \citep{PadoanNordlund2002,HennebelleChabrier2008,Hopkins2012b}. However, the driving mechanisms of this turbulence are not well understood and remain an active area of research.

Previous theoretical and numerical work has shown that the standard deviation of density fluctuations is linked to the sonic Mach number of the turbulence. The standard deviation of density fluctuations $\sigma_{\rho/\langle\rho\rangle}$ relative to the mean density $\langle\rho\rangle$ is related to the turbulent Mach number, $\mathcal{M} \equiv \sigma_v/c_s$, where $\sigma_v$ and $c_s$ are the standard deviation of the turbulent velocity fluctuations and the sound speed, through \citep{PadoanNordlundJones1997,FederrathKlessenSchmidt2008,KonstandinEtAl2012ApJ},
\begin{equation} \label{eq:sigrho-mach}
    \sigma_{\rho/\langle\rho\rangle} = b \mathcal{M}.
\end{equation}
The proportionality constant $b$ is known as the turbulence driving parameter, and its value depends on the modes induced by the turbulence driving mechanism \citep{FederrathKlessenSchmidt2008}. A purely solenoidal driving corresponds to $b=1/3$, whereas a purely compressive driving gives $b=1$, with $b$ smoothly increasing from $1/3$ to $1$ as the driving becomes progressively more compressive \citep{FederrathDuvalKlessenSchmidtMacLow2010}. Along with the density and velocity probability density functions (PDFs), the driving parameter $b$ therefore provides information on the mode of turbulence driving in different environments. For example, observations in \citet{PadoanNordlundJones1997}, \citet{Brunt2010}, and \citet{GinsburgFederrathDarling2013} measured column density and velocity fluctuations in molecular clouds in the Solar neighbourhood. With appropriate methods to estimate the volume density fluctuations from the column density fluctuations \citep{BruntFederrathPrice2010a,BruntFederrathPrice2010b,BruntFederrath2014,KainulainenFederrathHenning2014}, they estimated the $b$ parameter and typically found values $b\gtrsim0.5$ in the molecular clouds IC5146, Taurus, and GRSMC43.30-0.33, located in the spiral arms of the Milky Way, indicating primarily compressive driving of turbulence in these clouds. Such driving may be the result of supernova explosions, shock compression in the spiral arms, gravitational collapse, or radiation feedback from massive stars. The latter was confirmed by \citet{MenonEtAl2021} in several gas/dust pillars in the Carina Nebula, where they find \mbox{$b\sim0.7$--$1.0$}, i.e, the radiation from nearby massive stars that sculpted these pillars also drives compressive turbulence inside them. In other regions of the galaxy, the turbulence driving is more solenoidal, such as in the Central Molecular Zone (CMZ) cloud `Brick' (G0.253+0.016), where \citet{FederrathEtAl2016} found \mbox{$b\sim0.2$--$0.3$}. This predominantly solenoidal driving mode of turbulence is due to the strong shearing motions in the CMZ, driving turbulence in the clouds near the Galactic Centre. Thus, we expect different physical drivers of turbulence in various galactic environments to produce different values of $b$ \citep{FederrathEtAl2017iaus}.

The turbulence driving parameter $b$ plays a key role for star formation. Compressive driving ($b\sim1$) can produce star formation rates more than an order magnitude higher than solenoidal driving ($b\sim0.3$) \citep{FederrathKlessen2012,Federrath2018}. Thus, it is crucial to determine the driving mode $b$ of different physical drivers of turbulence, in order to understand and predict the star formation activity in different environments.

Here we aim to determine the driving mode of turbulence induced by hydrodynamical shocks. Such shock-driven turbulence is very common in the interstellar medium (ISM). For example, supernova (SN) explosions or radiation from massive stars drive shocks in the ISM, which drives turbulence \citep{MacLowKlessen2004}. Simulations of SN-driven turbulence find that the turbulence that the SN explosions produce is relatively solenoidal in the velocity field \citep{PanEtAl2016}. However, even purely compressive driving typically produces a $\sim50\%$ fraction of solenoidal modes in the velocity field \citep{FederrathDuvalKlessenSchmidtMacLow2010}, by non-linear shock interactions and baroclinic vorticity generation in a multi-phase (multi-temperature) medium \citep{DelSordoBrandenburg2011}. The presence of solenoidal modes in the velocity field, however, does not imply a dominance of solenoidal driving of the turbulence, and in fact, the density fluctuations may still follow Eq.~(\ref{eq:sigrho-mach}) with $b\sim1$. This is particularly true in a medium of nearly constant temperature, where the baroclinic term is ineffective, such as in dense molecular clouds, where cooling is very efficient and the gas remains roughly isothermal \citep[e.g., the pillars in the Carina nebula, studied by][]{MenonEtAl2021}. Further examples of flow driven turbulence include \citet{Giacalone&Jokipii2007}, \citet{GuoEtAl2012}, and \citet{Drury&Downes2012}, in which turbulence is generated through the differential acceleration of the upstream ISM, which occurs as a result of density inhomogeneities.

In order to determine the driving parameter $b$ of shock-driven turbulence, we run a set of numerical simulations, where we investigate the properties of the turbulence driven by a radiation-induced hydrodynamical shock impacting a pre-structured, multi-phase medium (modelled as a porous medium, i.e., foam). Our simulations are designed to mimic laboratory experiments of laser-induced, shock-driven turbulence, to be conducted at the National Ignition Facility (NIF), Lawrence Livermore National Laboratory. While these simulations provide predictions and calibrations for the laboratory experiments, they also provide the opportunity to study the statistics of shock-driven turbulence in relatively simple, controlled numerical experiments.

The paper is organised as follows. In Section~\ref{sec: methods} we present the basic geometry of the numerical and experimental setup, the hydrodynamical equations solved, and the numerical methods used here. In Section~\ref{sec:Results and Discussion} we discuss the density and velocity PDFs of the turbulence. We also discuss the thermodynamic properties of the multi-phase medium. We then study the dependence of the turbulence properties on the size of the structures in the foam (Section~\ref{sec: void comparison}), by considering three simulations with different foam void diameters. Using these results, we calculate the turbulence driving parameter of the shock-driven turbulence in Section~\ref{sec: driving parameter}. We also check if the driving parameter depends on the choice of analysis volume, time, or numerical resolution (Appendices). We compare our results with those of astrophysical observations and other simulations in Section~\ref{sec: final}. Section~\ref{sec: summary} summarises the results and conclusions of this study.

\section{Methods}
\label{sec: methods}

\subsection{Setup and geometry} \label{sec:setup}
\begin{figure}
    \includegraphics[width=1.0\linewidth]{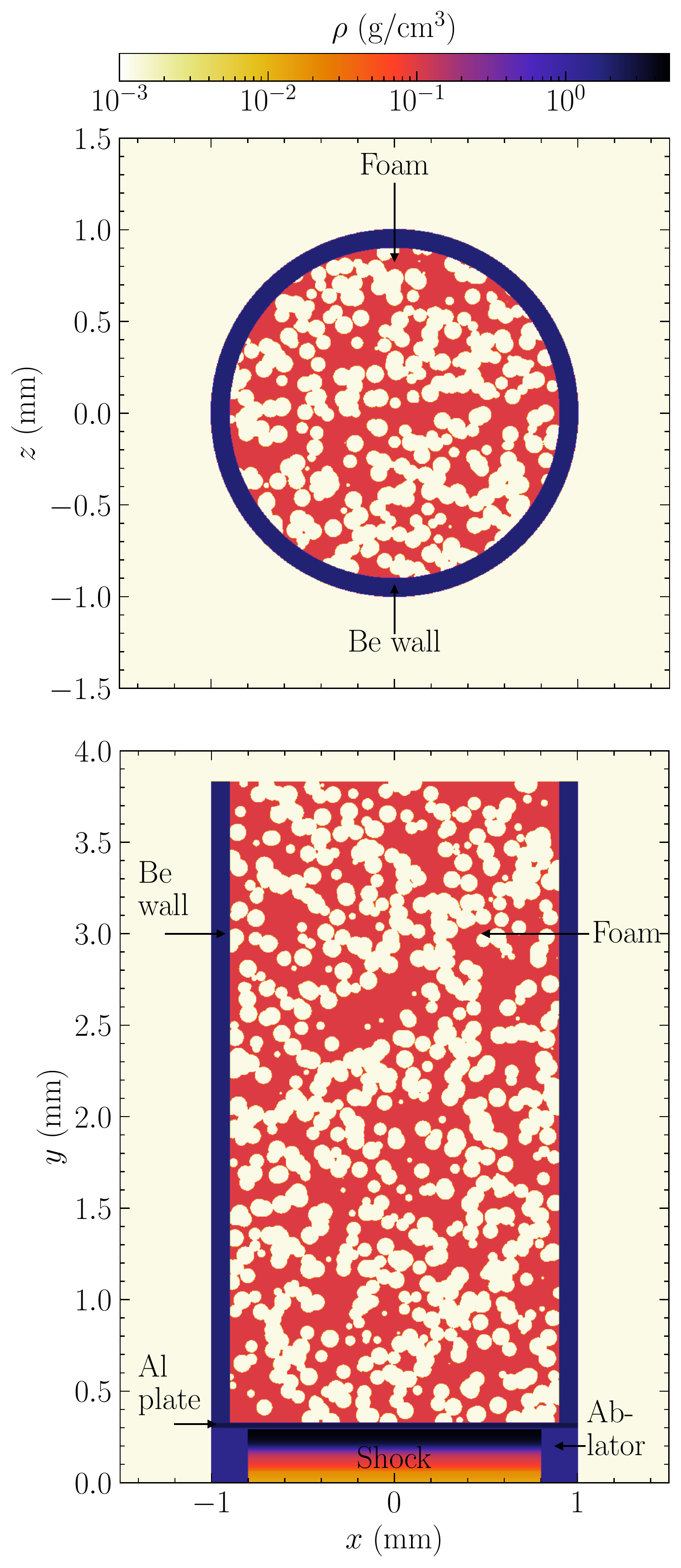}
    \caption{Volume density slices at time $t=0$ (initial conditions), showing the geometry of the basic setup. The top panel shows the $xz$-slice of the material (volume) density taken at $y=2\,\mathrm{mm}$, and the bottom panel shows a respective $xy$-slice taken at $z=0\,\mathrm{mm}$. The hydrodynamical shock wave is initialised in the plastic ablator material ($\rho_0=1.4\,\mathrm{g/cm^3}$), seen in the bottom panel, and then travels into the foam (here shown with void diameter of $50\,\mu\mathrm{m}$). The circular white structures represent voids, meant to roughly correspond to the cell structure of an actual foam. The non-void material (red) is a CH-based polymer with a density of $\rho_0 = 0.1\,\mathrm{g/cm^3}$.}
    \label{fig:initial_conditions}
\end{figure}
We consider a cylindrical tube with an outer diameter of $2.0\,\mathrm{mm}$ and a total length of $3.8\,\mathrm{mm}$ as shown in Fig.~\ref{fig:initial_conditions}. The walls of the tube with a thickness of $0.1\,\mathrm{mm}$ are made of beryllium, which has a density of 1.85~g/cm$^3$. The tube is filled with foam whose void diameter is $50\,\mu\mathrm{m}$; however, we are also considering cases with smaller and bigger void sizes below. The foam has an adiabatic index $\gamma=5/3$. At the bottom of the tube is a $0.3\,\mathrm{mm}$ thick plastic (CH-based) ablator, on which the laser beam is incident. A thin aluminium plate is placed on top of the ablator. The density of the ablator is 1.4~g/cm$^3$ and that of the aluminium plate is 2.7~g/cm$^3$. The details of this setup, including the specific materials used, are only important for the laboratory experiment, however, these specifics will likely change during the current design phase of the experiment, most notably the details of the shock tube \citep{NagelEtAl2017} and foam material \citep{HamiltonEtAl2016}. However, none of these specifics and details are important for the numerical simulations studied here, as we will only extract information in the foam material sufficiently far away from the boundaries, such that we can study the statistics of the turbulence in the foam, without any influence from the cylinder walls.

\subsection{Initial and boundary conditions} \label{sec:ics}
All materials are in pressure equilibrium at the start of the simulation. The laser is directed at the tube from the bottom (incident from the negative $y$ direction) and induces a hydrodynamic shock wave, which is initialised in the ablator material. As an initial condition for this hydrodynamical shock wave, we use a detailed radiation-hydrodynamical simulation of a 1D shock (see details in Appendix~\ref{sec:1dshock}). We then interpolate this 1D shock into the 3D domain, spanning the whole width (in $x$ and $z$ of the ablator). This interpolation serves to initialise a plane-parallel shock travelling in the $y$ direction, starting in the ablator at $y=0.3\,\mathrm{mm}$.

The assumed foam material (excluding voids) in these simulations has a density of $0.1\,\mathrm{g/cm^3}$. The mean density of the foam (including the voids) is set to $0.05\,\mathrm{g/cm^3}$. The voids are modelled as spheres with a diameter of $50\,\mathrm{\mu m}$ for the standard simulation shown in Fig.~\ref{fig:initial_conditions}, and have the density of air ($1.225\times10^{-3}\,\mathrm{g/cm^3}$). An increasing number of voids is placed randomly (sampled from a uniform distribution of positions) in the foam, until the total mean density of the foam has reached the target mean density of $0.05\,\mathrm{g/cm^3}$. The voids are allowed to overlap.

The simulations use outflow boundary conditions, i.e., once the shock breaks out of the cylinder, it is free to leave the computational domain. The same applies in all spatial directions, i.e., ultimately, the wall materials will also leave the computational domain. This means that there is no influence of the numerical boundary on the results, because we are only extracting information from within a small portion of the domain, where the medium can be considered turbulent, and no information is allowed to propagate into the domain from the boundaries.

\subsection{Simulation methods}
We use a modified version of the FLASH code \citep{FryxellEtAl2000,DubeyEtAl2008} (v4), to solve the three-dimensional compressible hydrodynamical equations, with the setup and initial conditions specified in Sec.~\ref{sec:setup} and~\ref{sec:ics}. The system of hydrodynamical equations is solved with the robust HLL5R approximate Riemann scheme \citep{WaaganFederrathKlingenberg2011}. We also compared the present simulations using the HLL5R solver with the PPM solver \citep{ColellaWoodward1984}, and find very good agreement.

The computational grid has $768\times1024\times768$ compute cells, capturing the large-scale turbulent dynamics and sufficient to converge on the statistical properties of the turbulent gas \citep{KitsionasEtAl2009,PriceFederrath2010,KritsukEtAl2011Codes}. This is also confirmed by a resolution study presented in Appendix~\ref{sec:resolution}. The number of cells in the $y$-direction is larger than in the $x$ and $z$ directions, which exactly compensates for the different computational domain length in $y$ as opposed to $x$ and $z$. Therefore, the physical cell width is the same in all three directions. We are not using any explicit viscosity in these simulations, i.e., we are conducting Implicit Large Eddy Simulations (ILES). Thus, the Reynolds number is not directly controlled, but increases with the numerical grid resolution. Considering a resolution of $N$ cells in linear direction across the turbulent region, grid codes in ILES mode usually result in Reynolds numbers of the order of $\mathrm{Re}\sim N^{4/3}$ \citep{BenziEtAl2008,FederrathEtAl2011,McKeeEtAl2020}, which means that we expect to develop turbulent flow in the simulations, i.e., typically $\mathrm{Re}>10^3$ \citep{Frisch1995}. In the real experiments, we also expect turbulent flow \citep{RobeyEtAl2003}.

For the thermodynamics, we use an ideal equation of state (EOS), where all materials have an adiabatic index $\gamma=5/3$. The foam is modelled as ionised CH-based polymer, which has a mean molecular weight of 6.5, and the foam voids are assumed to be filled with air at a mean molecular weight of 29. This is a simplified approach, where phase transitions, chemistry, radiation and cooling of the materials are neglected. While these processes would introduce some quantitative changes, e.g., in the details of the shock Mach number, temperature of the foam, and mean molecular weight during the shock passage, the qualitative evolution remains the same. In particular, for the purposes of understanding the turbulence driving mode of this system, the details of the thermodynamical modelling do not affect our basic results and conclusions. This is because the driving parameter is ultimately a property of the shock driving itself, and the response of the system being shocked is such that we are only interested in the ratio of relative density dispersion $\sigma_{\rho/\langle\rho\rangle}$ and turbulent Mach number; see Eq.~(\ref{eq:sigrho-mach}). Thus, while the details of the thermodynamics, chemistry, radiation and cooling will alter the density dispersion and Mach number in their absolute values (for instance, strong cooling would allow for higher $\mathcal{M}$), the ratio of these dimensionless numbers, i.e., the driving parameter $b$, is expected to remain largely invariant.

For the same reason, we can apply the basic findings from these numerical experiments to shock-driven turbulence in the ISM. While the absolute values of temperature, density, etc., are very different in the ISM compared to the laboratory experiment, the driving parameter only depends on dimensionless numbers that characterise the turbulent flow, which can be compared for systems of vastly different absolute scale, because turbulence is a ubiquitous, scale-free process.

\subsection{Analysis methods}
Our goal is to study the properties of shock-driven turbulence in a pre-structured medium. Here we measure the density and velocity distributions in the turbulent shocked foam, and finally obtain the turbulence driving parameter $b$.

\begin{figure*}
	\includegraphics[width=\linewidth]{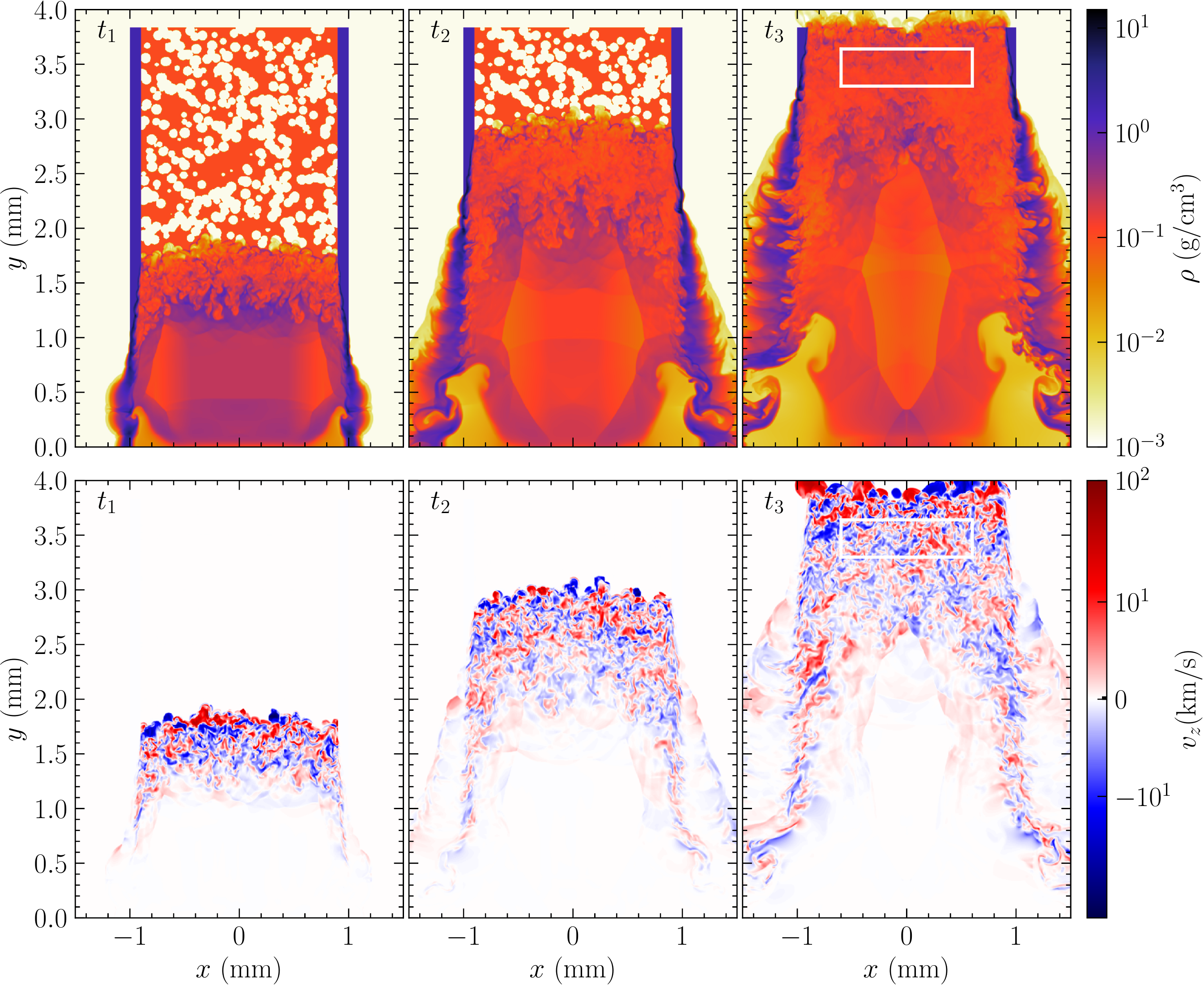}
    \caption{Time evolution of the simulation with $50\,\mathrm{\mu m}$ foam voids, showing $xy$-slices of the material density (top panels) and the $z$-component of the velocity (bottom panels) at $z=0$. The snapshots are taken at times $t_1=25\,$ns, $t_2=50\,$ns and $t_3=75\,$ns (from left to right). The white rectangles in the right-hand panels show the region in which we carry out the subsequent turbulence analysis.}
    \label{fig:time_evolution}
\end{figure*}

We first consider a foam with void size of $50\,\mathrm{\mu m}$. In order to study the properties of the turbulence, we select a time after the initiation of the shock, at which there is a large enough region of well-developed turbulence. A cuboidal region of sufficiently developed turbulence is selected at time $t=75\,\mathrm{ns}$, which is represented by the white rectangle in Fig.~\ref{fig:time_evolution} at $t=t_3$. The analysis region has dimensions of $1.2\,\mathrm{mm} \times 0.34\,\mathrm{mm} \times 1.2\,\mathrm{mm}$. The density, sound speed and Mach number PDFs are obtained and analysed for this region, and the turbulence driving parameter is calculated.

We also investigate the dependence of the turbulence properties on the foam void size by considering two more simulations with a smaller ($12.5\,\mathrm{\mu m}$) and a larger ($100\,\mathrm{\mu m}$) foam void size along with the standard $50\,\mathrm{\mu m}$ foam void case. The analysis procedure is the same as that for the $50\,\mathrm{\mu m}$ void case. The time evolution and volume dependence of the turbulence is also studied by considering different time instances and different analysis volumes for the $50\,\mathrm{\mu m}$ foam void size case (Appendices~\ref{sec: volume dependence} and~\ref{sec: time dependence}).

\section{Results and Discussion}
\label{sec:Results and Discussion}

\subsection{Time evolution and structure of laser-induced turbulence}

\begin{table}
    \centering
    \caption{Physical parameters of the turbulence for the foam with $50\,\mathrm{\mu m}$ voids.}
    \def\arraystretch{1.0}
    \setlength{\tabcolsep}{8pt}
    \begin{tabular}{ccc}
    \hline
    Quantity & Symbol & Value\\
    \hline
    Mean density & $\langle \rho \rangle$ & $0.15\,$g/cm$^3$ \\
    Mean sound speed & $\langle c_s \rangle$ & $20\,$km/s\\
    Mean velocity along $x$-direction & $\langle v_x \rangle$ & $-0.05\,$km/s\\
    Mean velocity along $y$-direction & $\langle v_y \rangle$ & $25\,$km/s\\
    Mean velocity along $z$-direction & $\langle v_z \rangle$ & $-0.04\,$km/s\\
    Mean temperature & $\langle T \rangle$ & $2.1\times10^5\,$K\\
    \hline
    \end{tabular}
    \label{physical_params}
\end{table}

Fig.~\ref{fig:time_evolution} depicts the time evolution of the shock-driven turbulence in the foam with void size of $50\,\mathrm{\mu m}$. As the shock passes through the foam from the bottom, it drives turbulent density and velocity fluctuations (see top and bottom panels of Fig.~\ref{fig:time_evolution}, respectively). As the shock wave propagates into the foam tube, the walls melt, and are pushed outwards. The volume of the turbulent region increases as the shock progresses further. Finally, the shocked foam bursts out of the tube at the top ($y=3.8\,\mathrm{mm}$). The whole process happens within $\sim100\mathrm{ns}$.

The right-hand panels at $t=t_3$ of Fig.~\ref{fig:time_evolution} show the time at which we carry out our analysis. The cuboidal region selected for the turbulence analysis is shown as a white rectangle. This region was selected to minimise the impact of systematic biases from the boundaries or the shock passing through the foam. In Appendix~\ref{sec: volume dependence} and~\ref{sec: time dependence}, we show that our main results do not depend on the specific choice of the analysis region or the time of the analysis.

\begin{figure*}
    \includegraphics[width=\linewidth]{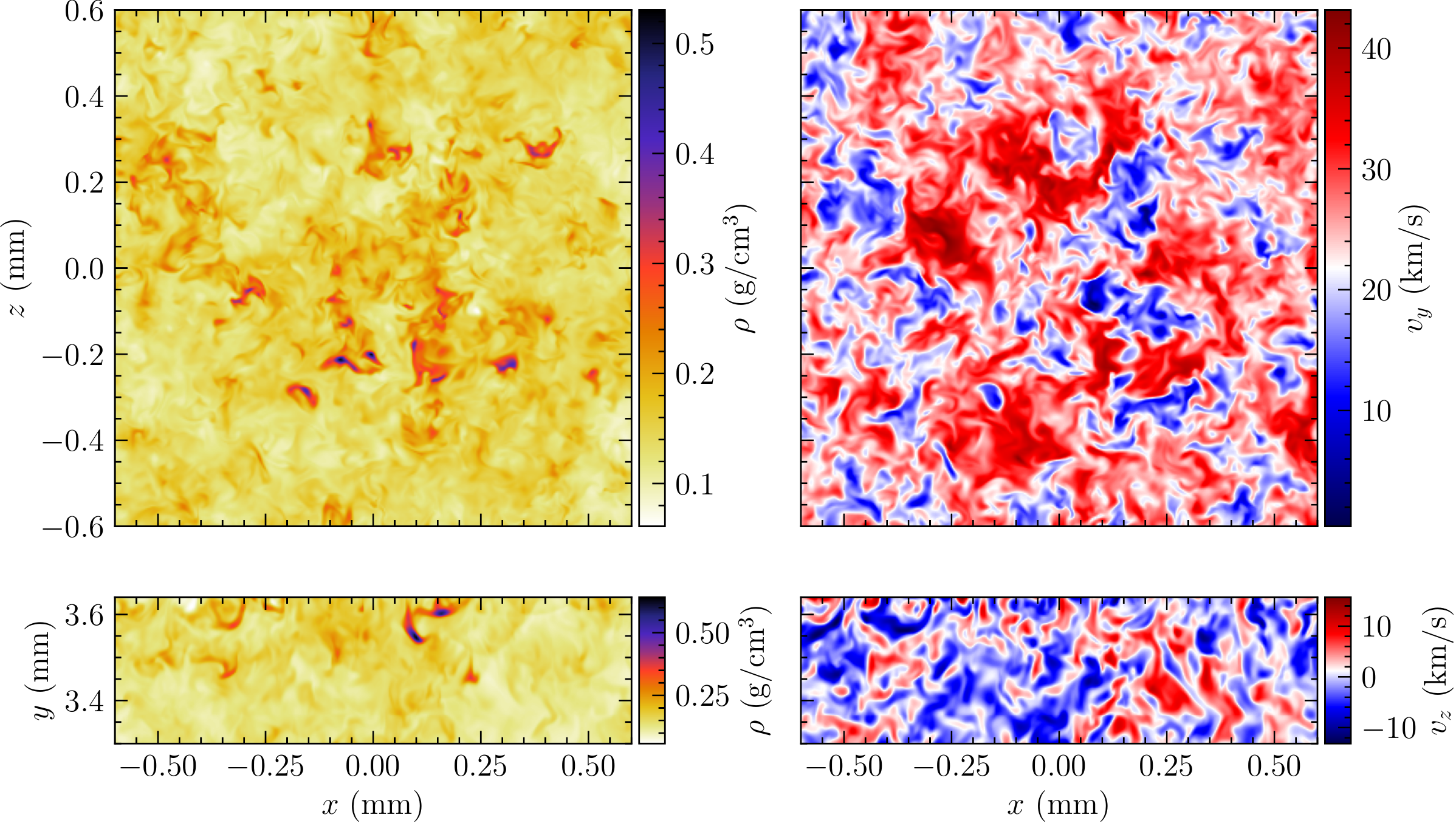}
    \caption{Maps showing the turbulence analysis region in an $xz$ slice of the density (top left panel), an $xz$ slice of the $y$-velocity (top right panel), an $xy$ slice of the density (bottom left panel), and an $xy$ slice of the $z$-velocity (bottom right panel). All the slices pass approximately through the centre of the analysis region, at $(x,y,z) = (0.0,3.5,0.0)\,$mm. As seen from the velocity slices, there is a net velocity in the $y$-direction, along which the shock propagates. Despite this non-zero bulk speed, the density and velocity structures are largely isotropic, and exhibit a morphology typical of fully-developed turbulence.}
    \label{fig: analysis region}
\end{figure*}

Fig.~\ref{fig: analysis region} shows the density (left) and velocity (right) snapshots of the analysis region. The component of the velocity perpendicular to the plane of the slice is shown. It is seen that there is a net velocity along the direction of the propagation of the shock, i.e., the shock induces a systematic bulk motion along the $y$-direction. The average velocities in the $x$ and $z$ directions are very close to zero. The average physical properties of the analysis region are listed in Table~\ref{physical_params}. The small bulk velocities in the $x$ and $z$ directions are because of density inhomogeneities and because the analysis region does not cover the entire simulation volume\footnote{The total $x$ and $z$ momentum in the simulation domain is initially zero, and remains zero until material leaves through the simulation boundaries.}. However, these bulk velocities do not affect the results of this study, as we subtract them in the definition of the local Mach number, which is the quantity of interest. Despite the non-zero bulk flow along the $y$ direction, the system is largely isotropic. The velocity fluctuations along the shock direction are somewhat larger than perpendicular to it, which we will quantify in detail below. The density fluctuations are relatively independent of the direction. In addition to these bulk flows, we also find that there are weak density and velocity gradients. In Appendix~\ref{sec: gradients} we investigate the influence of these gradients, and find that they are small compared to the main fluctuations in the velocity and density. Thus, our following analyses are practically independent of the presence of these systematic gradients.

\subsection{Density and sound speed distributions}
\label{sec: rho_cs_PDF}

\begin{figure}
\centering
\includegraphics[width=\columnwidth]{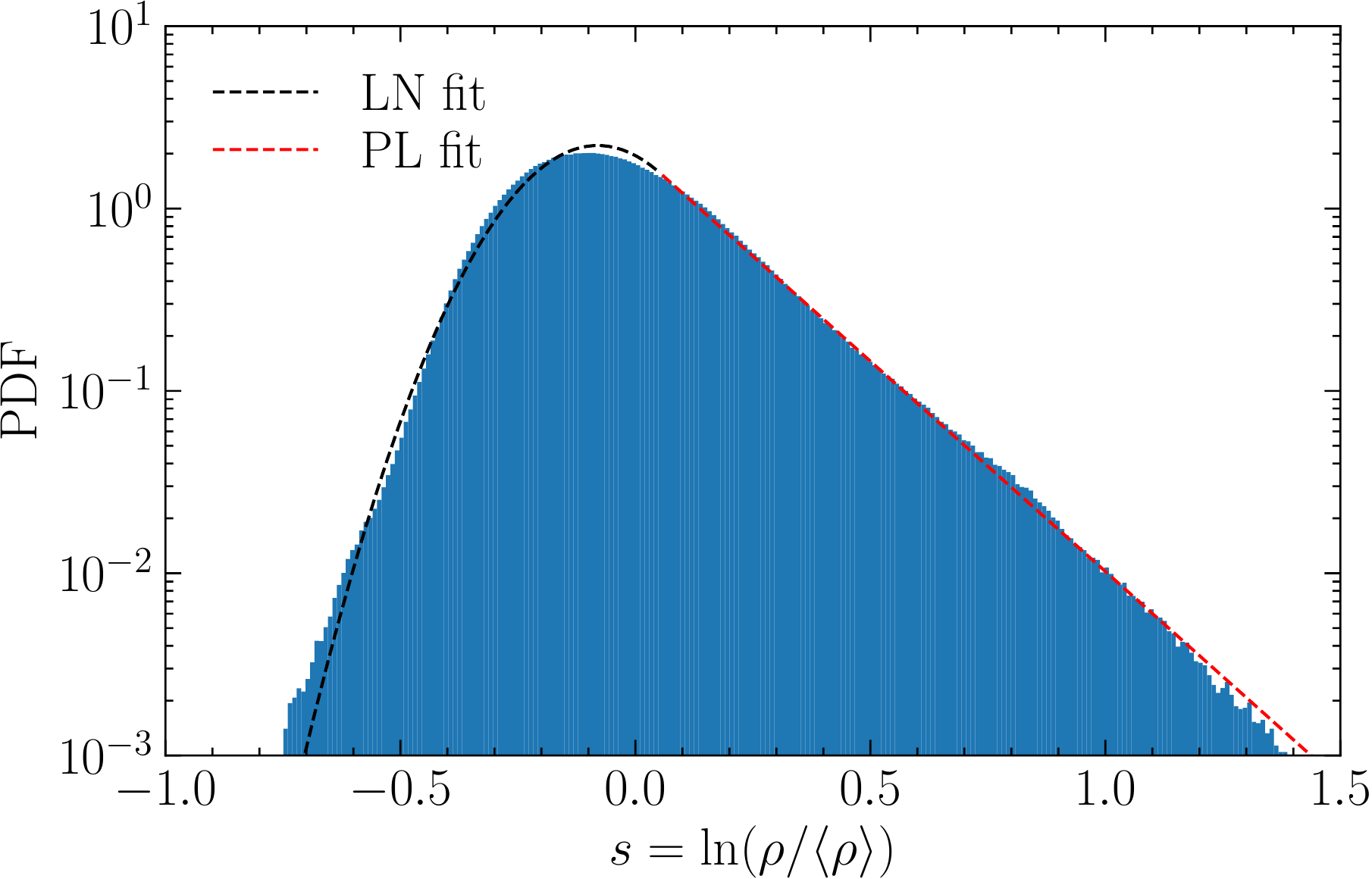}
\caption{Probability density function (PDF) of the logarithmic density contrast $s=\ln(\rho/\langle\rho\rangle)$ in the analysis region. The PDF follows a Gaussian distribution at low $s$ up to around the peak, followed by a power-law tail towards high densities. The dashed lines represent the log-normal plus power-law (LN+PL) fit via Eq.~(\ref{LNPL_eq}), with fit parameters reported in Tab.~\ref{table:fit}.}
\label{sPDF_50}
\end{figure}

\begin{figure}
    \centering
    \includegraphics[width=\columnwidth]{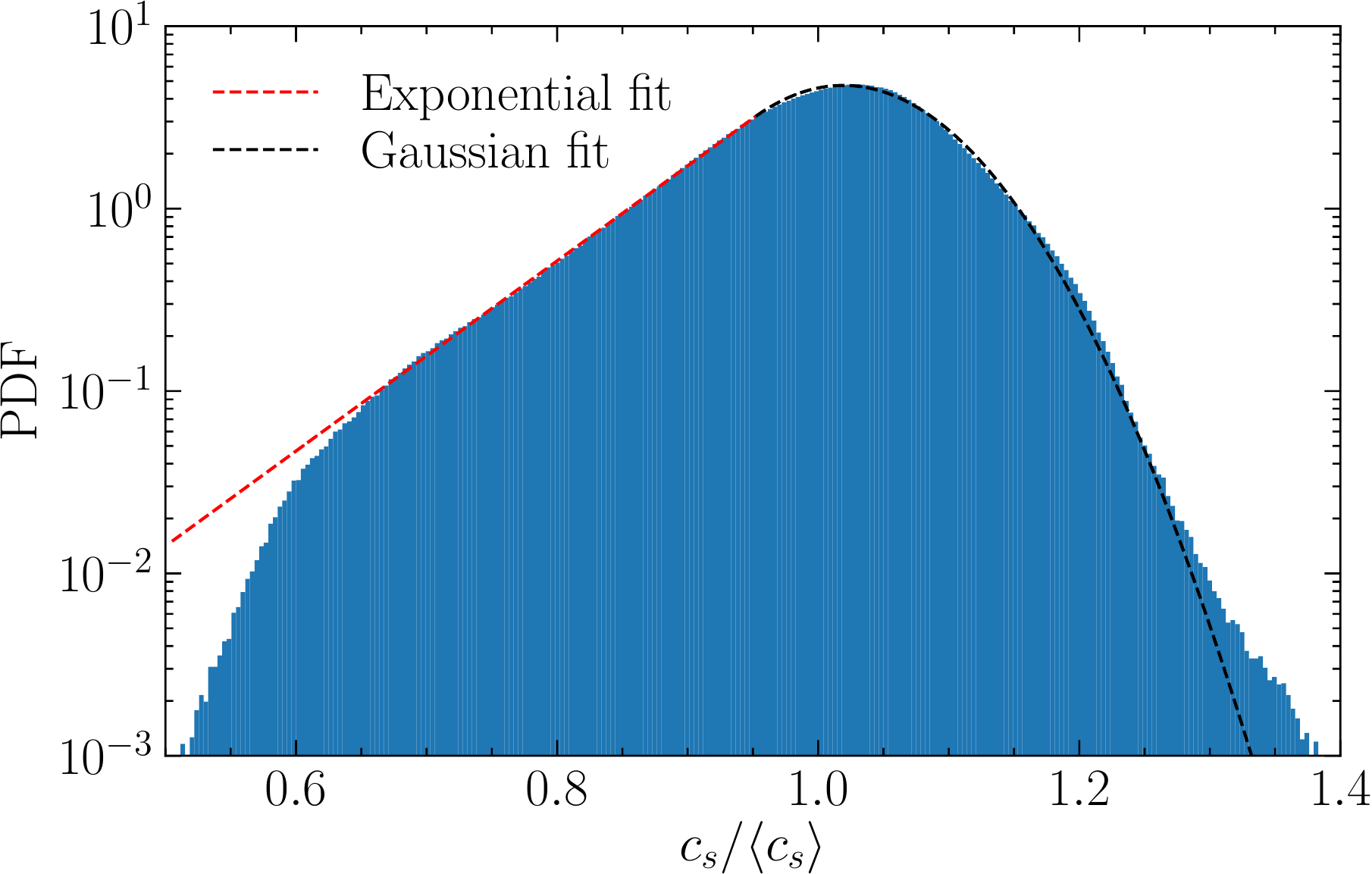}
    \caption{Sound speed PDF of the gas in the analysis region, where sound speed is defined as $c_s = \sqrt{\gamma p_\mathrm{th}/\rho}$ with the mean sound speed $\langle c_s \rangle$. The dashed lines represent an Exponential + Gaussian fit (see Eq.~\ref{cs_fit}), the details of which are provided in Appendix~\ref{fit_appendix}.}
    \label{cs_PDF}
\end{figure}

\begin{figure*}
    \centering
    \includegraphics[width=\linewidth]{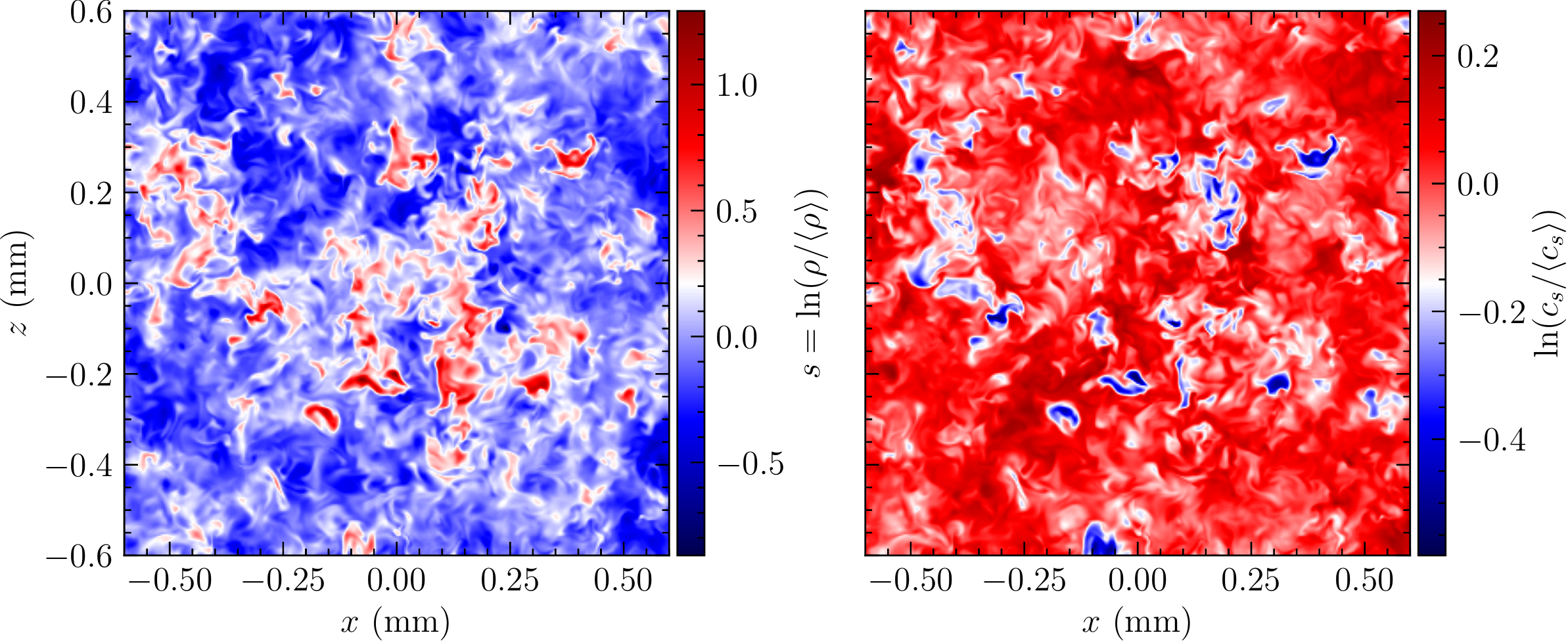}
    \caption{Maps showing $xz$-slices of the density contrast parameter, $s$ (left panel), and sound speed (right panel) taken at $y=3.5\,$mm. Both quantities are scaled with respect to their respective mean values in the analysis region, and the natural logarithm is taken for better visualisation and comparison. We see that regions of higher density correspond to lower sound speed and vice versa.}
    \label{density_cs}
\end{figure*}

Here we quantify the density fluctuations and thermodynamic properties of the turbulence in the analysis region.

\subsubsection{Gas density distribution}

In order to quantify the properties of the turbulent gas density, we study the probability density function (PDF) of the logarithmic density contrast,
\begin{equation}
s \equiv \mathrm{ln}(\rho/\langle \rho \rangle),    
\end{equation}
with the mean density $\langle \rho \rangle$ in the analysis region. The quantity $s$ has long been used in the literature to study the density PDF, because in this transformation to the natural logarithm of $\rho/\langle \rho \rangle$, the PDF of $\rho$ is often found to be nearly log-normal for isothermal turbulence, that is, the PDF of $s$ is Gaussian \citep{Vazquez1994,PassotVazquez1998,KritsukEtAl2007,FederrathKlessenSchmidt2008,FederrathDuvalKlessenSchmidtMacLow2010}.

Fig.~\ref{sPDF_50} shows the PDF of $s$ in our turbulence analysis region. We see that the PDF resembles a Gaussian distribution at lower densities, and exhibits a high-density power-law tail. We model this density PDF as a piece-wise continuous log-normal + power-law (LN+PL) function\footnote{The wording `log-normal + power-law (LN+PL)' strictly-speaking only applies to the PDF of $\rho$. When transformed to $s$, this means that the PDF of $s$ is a Gaussian + exponential. However, for simplicity, we will generally refer to this form of the density PDF as LN+PL, even when considering the quantity $s$.}, following the functional form used by \citet{CollinsEtAl2012}, \citet{Burkhart2018} and \citet{KhullarEtAl2021} for the density PDFs,
\begin{equation}
    p(s)=
    \begin{cases}
    \frac{N}{\sqrt{2 \pi \sigma_s^2}}\exp\left[-\frac{(s-s_0)^2}{2\sigma_s^2}\right] \qquad &\text{ for }s < s_t,\\ Np_0\exp{(-\alpha s)} \qquad &\text{ for } s\ge s_t.
\end{cases} 
\label{LNPL_eq}
\end{equation}
Here, $N$ and $p_0$ are normalisation factors, and $\alpha$ is the slope of the power-law tail. The parameters $\sigma_s$ and $s_0$ are the standard deviation and the mean of the LN part, respectively. We impose three constraints on this functional form:
\begin{enumerate}
    \item The PDF is normalised, such that $\int p(s)ds = 1$.
    \item The PDF is continuous at the transition point, $s_t$.
    \item The derivative of the PDF, $dp(s)/ds$, is continuous at $s_t$.
\end{enumerate}
Because of these constraints, only 3 out of the 6~parameters in Eq.~(\ref{LNPL_eq}) are independent. We choose the slope of the PL ($\alpha$), the width of the LN part ($\sigma_s$), and the transition point ($s_t$) as the free parameters. The remaining 3~parameters, $N$, $p_0$, and $s_0$, are determined from the 3~free parameters and the imposed 3~constraints.

We fit Eq.~(\ref{LNPL_eq}) to the PDF data. As seen from Fig.~\ref{sPDF_50}, the LN+PL PDF provides a very good fit. The parameters of the LN+PL fit are provided in Table~\ref{table:fit}.

\subsubsection{The origin of the density PDF power-law tail}

Log-normal + power-law tails in the density PDF have been detected in observations of interstellar clouds \citep{KainulainenEtAl2009,SchneiderEtAl2011,SchneiderEtAl2013,KainulainenFederrathHenning2013,KainulainenFederrathHenning2014,KainulainenFederrath2017,AlvesEtAl2017}. The most common origin of these high-density PL tails is gravitational collapse \citep{Klessen2000,KritsukNormanWagner2011,FederrathKlessen2013,GirichidisEtAl2014,BurkhartMocz2019,KoertgenFederrathBanerjee2019,JaupartChabrier2020,KhullarEtAl2021}. A gravitational origin of the PL tail seen in our simulation is of course excluded. However, two other physical processes can also produce high-density PL tails: 1) strong shock compression \citep{TremblinEtAl2014}, and 2) an effective polytropic index $\Gamma<1$, in the equation of state for the thermal gas pressure, $p_\mathrm{th}\propto\rho^\Gamma$ \citep{PassotVazquez1998,FederrathBanerjee2015}.

\citet{TremblinEtAl2014} found that compression by expanding HII regions in molecular clouds can produce high-density power-law tails, similar to what is observed in our case. Considering that the compression induced by the hydrodynamical shock wave is significant, it is plausible that there is at least some contribution from the shock compression to creating the high-density PL tail in the PDF. An additional contributing factor to the PL tail is the thermodynamics and the multi-phase nature of the medium. The foam consists of a 2-phase medium, i.e., the dense foam cell wall material, and the low-density voids in between. Since the system is initially in pressure equilibrium, this means that the two phases have different temperature, i.e., the dense phase is colder than the low-density phase. This gives rise to a distribution of temperatures, and therefore, a distribution of sound speeds in the gas.

Fig.~\ref{cs_PDF} shows the PDF of the local sound speed, $c_s = \sqrt{\gamma p_\mathrm{th}/\rho}$, with the adiabatic index $\gamma=5/3$, used to model the materials in the simulations. While the variations in the sound speed are relatively small (overall variation by a factor $\sim2$), we clearly see an exponential tail in the sound speed PDF towards low values of $c_s$, corresponding to colder, denser gas, as opposed to the Gaussian component seen at higher $c_s$ (warmer, lower-density gas). The sound speed PDF therefore mirrors the shape of the density PDF, suggesting a link between the two.

To visualise this link, we show $xz$-slices of the density contrast parameter $s$ and the sound speed in Fig.~\ref{density_cs}. It is clearly seen that over-dense regions have lower sound speed and vice versa. This is because initially, owing to pressure equilibrium, the foam (which has a higher density) is cooler as compared to the air in the foam voids (which has a lower density). These two phases of the medium, i.e., the foam cell wall material and the air in the voids, mix as the shock passes and the medium becomes turbulent. Since the adiabatic $\gamma=5/3$, the compressed regions will heat up. However, the previously over-dense foam cell wall material still stays cooler compared to the under-dense voids.

\begin{figure}
    \centering
    \includegraphics[width=\columnwidth]{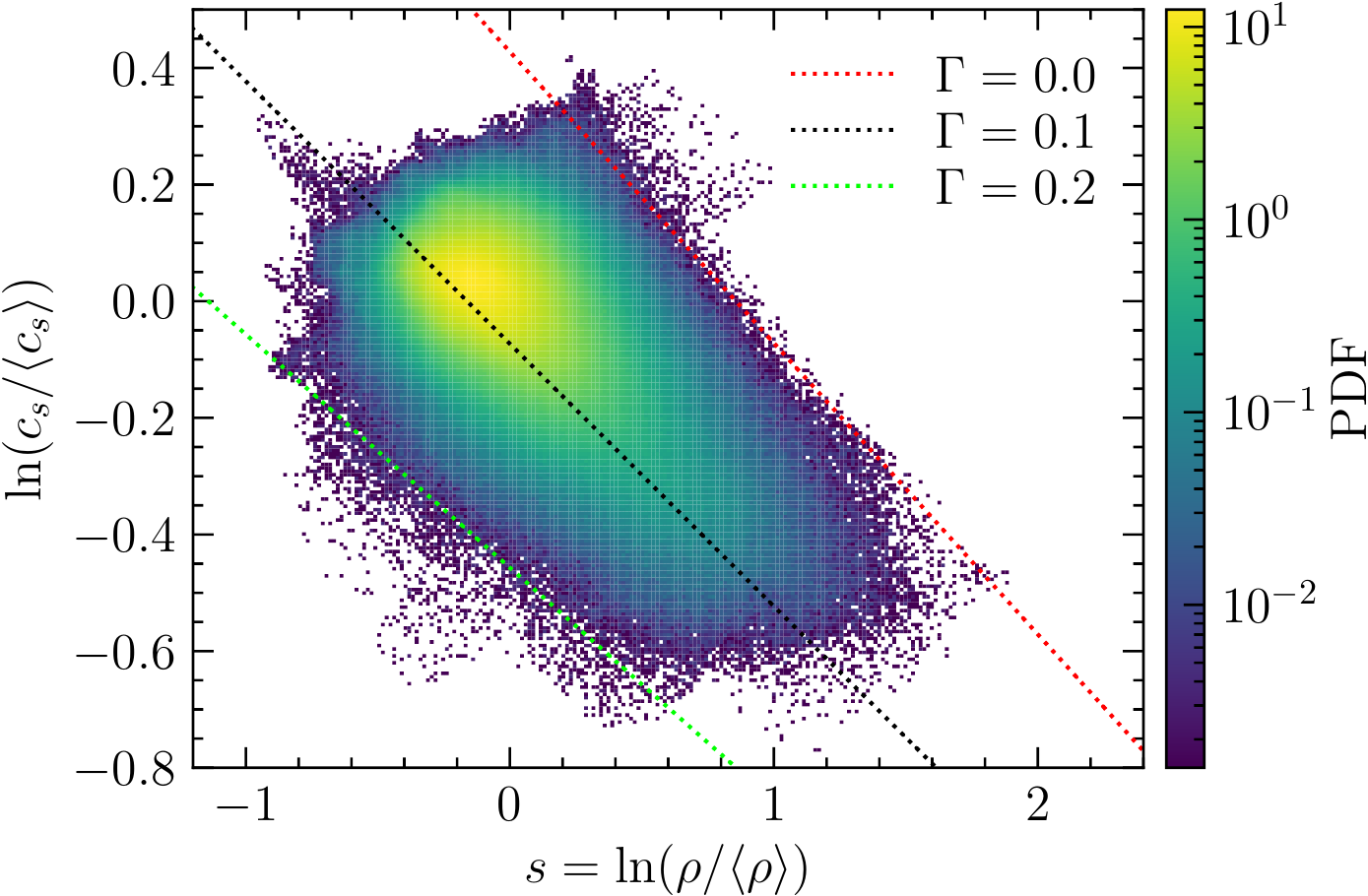}
    \caption{Density -- sound speed correlation PDF for the gas in the analysis region. We see a clear negative correlation between density and sound speed. For comparison and to guide the eye, three lines of constant polytropic $\Gamma$ are plotted, one each at the upper and the lower edge of the PDF, and one passing approximately through its centre.}
    \label{2dpdf}
\end{figure}

In order to quantify the link between density and sound speed, we show the density -- sound speed correlation PDF of the analysis region in Fig.~\ref{2dpdf}. The PDF indicates a negative correlation between the two quantities. This is in agreement with what is observed from the slice plots. Most of the points have relatively lower density and higher sound speed. However, there is a large tail corresponding to cooler, over-dense gas. For a polytropic gas, $c_s \propto (p_\mathrm{th}/\rho)^{1/2} \propto (\rho^{\Gamma}/\rho)^{1/2} \propto \rho^{(\Gamma-1)/2}$ \citep{FederrathBanerjee2015}. In the correlation PDF, we plot lines of constant polytropic $\Gamma$ to find the effective $\Gamma$ of the system. From Fig.~\ref{2dpdf}, we see that the system has an effectively soft equation of state with $\Gamma \sim 0.0$ to $0.2$. As demonstrated and quantified in \citet{PassotVazquez1998} and \citet{FederrathBanerjee2015}, a polytropic exponent of $\Gamma<1$ gives rise to a high-density power-law tail in the $s$-PDF and a low-$c_s$ power-law region in the sound speed PDF, which is what we observed in Figs.~\ref{sPDF_50} and~\ref{cs_PDF}, respectively. Thus, we conclude that the high-density tail in the density PDF and the low-$c_s$ tail in the sound speed PDF are the result of the multi-phase nature of the medium, with an effective polytropic $\Gamma<1$.

\subsubsection{Density fluctuations}

Now that we have gained a basic understanding of the features (LN+PL) in the density PDF, we can focus our attention on the density fluctuations, $\sigma_{\rho/\langle\rho\rangle}$ in the analysis region. The overall turbulent density fluctuations are required in order to determine the turbulent driving parameter $b$ via Eq.~(\ref{eq:sigrho-mach}).

Since the contribution of the turbulent density fluctuations in the power-law tail of the PDF is substantial, we cannot use the value of $\sigma_s$ obtained from the LN+PL fit for this purpose, as it only measures the density fluctuations in the LN part of the PDF. We therefore follow two separate approaches to determine $\sigma_{\rho/\langle\rho\rangle}$:
\begin{enumerate}
\item we measure $\sigma_{\rho/\langle\rho\rangle}$ directly from the simulation data (without using the fit), by summation over all computational cells in the analysis region (noting that all cells have the same volume),
\begin{equation}
\label{eq: sigma_rho formula}
\begin{split}
&\langle \rho \rangle = \frac{1}{n} \sum_{i=1}^n \rho_i,\\
&\sigma_{\rho/\langle\rho\rangle} = \left[\frac{1}{n} \sum_{i=1}^n \left(\rho_i/\langle\rho\rangle - 1\right)^2\right]^{1/2},
\end{split}
\end{equation}
where $n$ is the total number of cells in the analysis volume, and $\rho_i$ is the density in cell $i$.

\item we integrate over the fitted density PDF to compute the total density fluctuations as
\begin{equation}
\sigma_{\rho/\langle\rho\rangle} = \left[\int \left(\rho/\langle\rho\rangle-1\right)^2 p(s) ds\right]^{1/2},
\label{eq:pdfintegral}
\end{equation}
with $\rho/\langle\rho\rangle=\exp(s)$.

\end{enumerate}
The value of $\sigma_{\rho/\langle\rho\rangle}$ calculated directly from the data is $0.270$, whereas the value obtained by integrating over the fitted density PDF via Eq.~(\ref{eq:pdfintegral}) is $0.269$. Thus, the values of the total density fluctuations obtained using the two methods agree very well, with a deviation of only $\sim 0.4\%$.

\subsection{Mach number distributions}
\label{sec: mach}
\begin{figure*}
	\includegraphics[width=\linewidth]{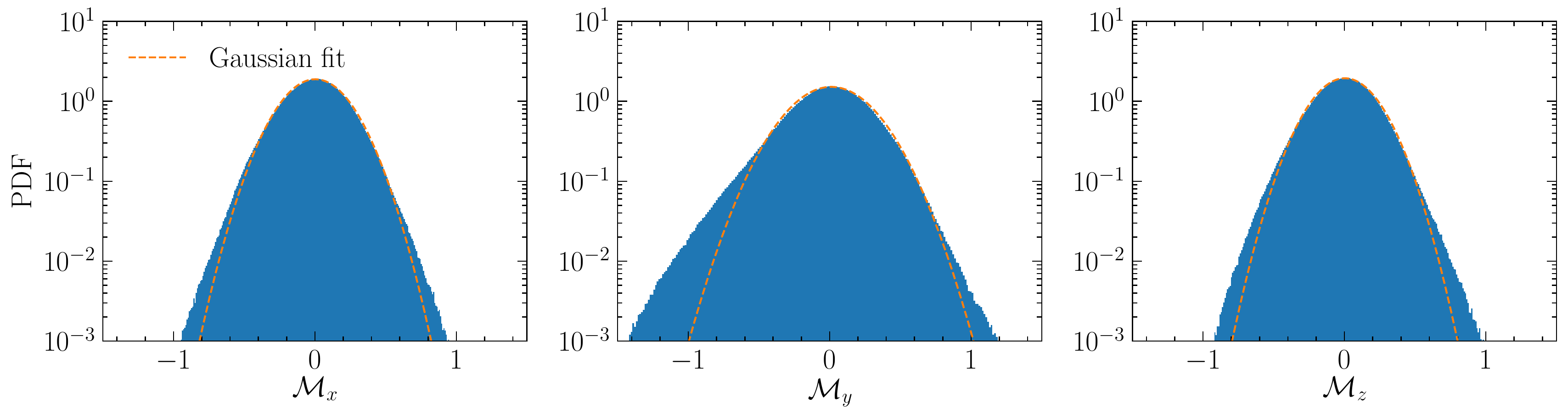}
    \caption{PDFs of the local Mach number in $x,\,y,\,z$ (from left to right) for shock-driven turbulence with $50 \,\mu$m foam voids. The Mach number PDFs are consistent with Gaussian fits via Eq.~(\ref{mach_gaussian}). The far wings of the PDFs have exponential tails, typical of intermittent fluctuations in the turbulence. However, these tails have no significant influence on the overall standard deviation of the distributions. We also see that the $\mathcal{M}_y$ PDF is slightly wider (by $\sim30\%$) than the $\mathcal{M}_x$ and $\mathcal{M}_z$ PDFs, due to $y$ being the shock propagation direction.}
    \label{fig:mach_PDFs}
\end{figure*}
Now that we have looked at the density PDF and calculated the standard deviation of the turbulent density fluctuations, we will focus our attention on the velocity statistics. In particular, we analyse the Mach number PDFs and obtain the 3D Mach number, which is required to finally calculate the turbulence driving parameter.
The local Mach number along the $x$-direction is defined as
\begin{equation}
    \mathcal{M}_x=\frac{v_x-\langle v_x \rangle}{c_s}
\end{equation}
where $c_s$ and $v_x$ are the sound speed and the $x$-component of the velocity at any given point, and $\langle v_x \rangle$ is the mean $x$-component of the velocity in the analysis volume. $\mathcal{M}_y$ and $\mathcal{M}_z$ are defined in analogy to $\mathcal{M}_x$.

Fig.~\ref{fig:mach_PDFs} shows the PDFs of $\mathcal{M}_x$, $\mathcal{M}_y$, and $\mathcal{M}_z$ in the analysis region. A Gaussian distribution is used to fit the Mach number distributions,
\begin{equation}
        p(\mathcal{M}_j)=\frac{1}{\sqrt{2 \pi \sigma_{\mathcal{M}_j}^2}}\exp\left[-\frac{\left(\mathcal{M}_j-\langle \mathcal{M}_j \rangle\right)^2}{2\sigma_{\mathcal{M}_j}^2}\right] \quad
        \text{with  } j \in \{ x,y,z \},
\label{mach_gaussian}
\end{equation}
where $\sigma_{\mathcal{M}_j}$ and $\langle \mathcal{M}_j \rangle$ are the standard deviation and the mean of the distribution, respectively.
As seen in Fig.~\ref{fig:mach_PDFs}, a Gaussian distribution provides a good fit for the Mach number PDFs, especially near the mean. This is typical of a purely turbulent medium, even at higher values of the Mach number \citep{Federrath2013}. The $\mathcal{M}_x$ and $\mathcal{M}_z$ distributions are almost identical, as a result of the $x-z$ symmetry of the setup. The $\mathcal{M}_y$ PDF is is somewhat wider, with $\sigma_{\mathcal{M}_y}$ about $30\%$ larger than $\sigma_{\mathcal{M}_x}$ or $\sigma_{\mathcal{M}_z}$. This is to be expected, as the shock progresses along the $y$-direction, inducing somewhat stronger velocity fluctuations along the shock direction. The Mach number PDFs have some non-Gaussian features, which manifest in the form of exponential wings on either side, far away from the mean. Such non-Gaussian wings are usually the result of intermittency in the turbulent flow \citep{SheLeveque1994,BoldyrevNordlundPadoan2002b,KritsukEtAl2007,SchmidtFederrathKlessen2008,FederrathDuvalKlessenSchmidtMacLow2010,KonstandinEtAl2012}. The wings in the PDF of $\mathcal{M}_y$ are asymmetrical, which is due to intermittency and/or the direct influence of the shock. The parameters of the Gaussian fits are listed in Table~\ref{table:fit}.

For comparison with the fitted values, we also compute the standard deviation of the Mach number directly from the data, in analogy to how we computed the standard deviation of the density fluctuations via Eq.~(\ref{eq: sigma_rho formula}). Finally, the 3D Mach number $\mathcal{M}$, which enters Eq.~(\ref{eq:sigrho-mach}), is calculated as
\begin{equation}
\mathcal{M}=\left(\sigma_{\mathcal{M}_x}^2+\sigma_{\mathcal{M}_y}^2+\sigma_{\mathcal{M}_z}^2\right)^{1/2}.
\label{eq:3dmach}
\end{equation}
Overall, we find that the turbulence behind the shock is subsonic. From the data, we find $\mathcal{M}=0.43$, while the Gaussian fits above give $\mathcal{M}=0.39$ (see Tab.~\ref{table:fit}), i.e., they are nearly identical, with the fitted value being $10\%$ smaller due to the fact that the fit underestimates the velocity fluctuations in the far wings of the distribution.

\section{Effects of different void size}
\label{sec: void comparison}

\begin{figure*}
	\includegraphics[width=\linewidth]{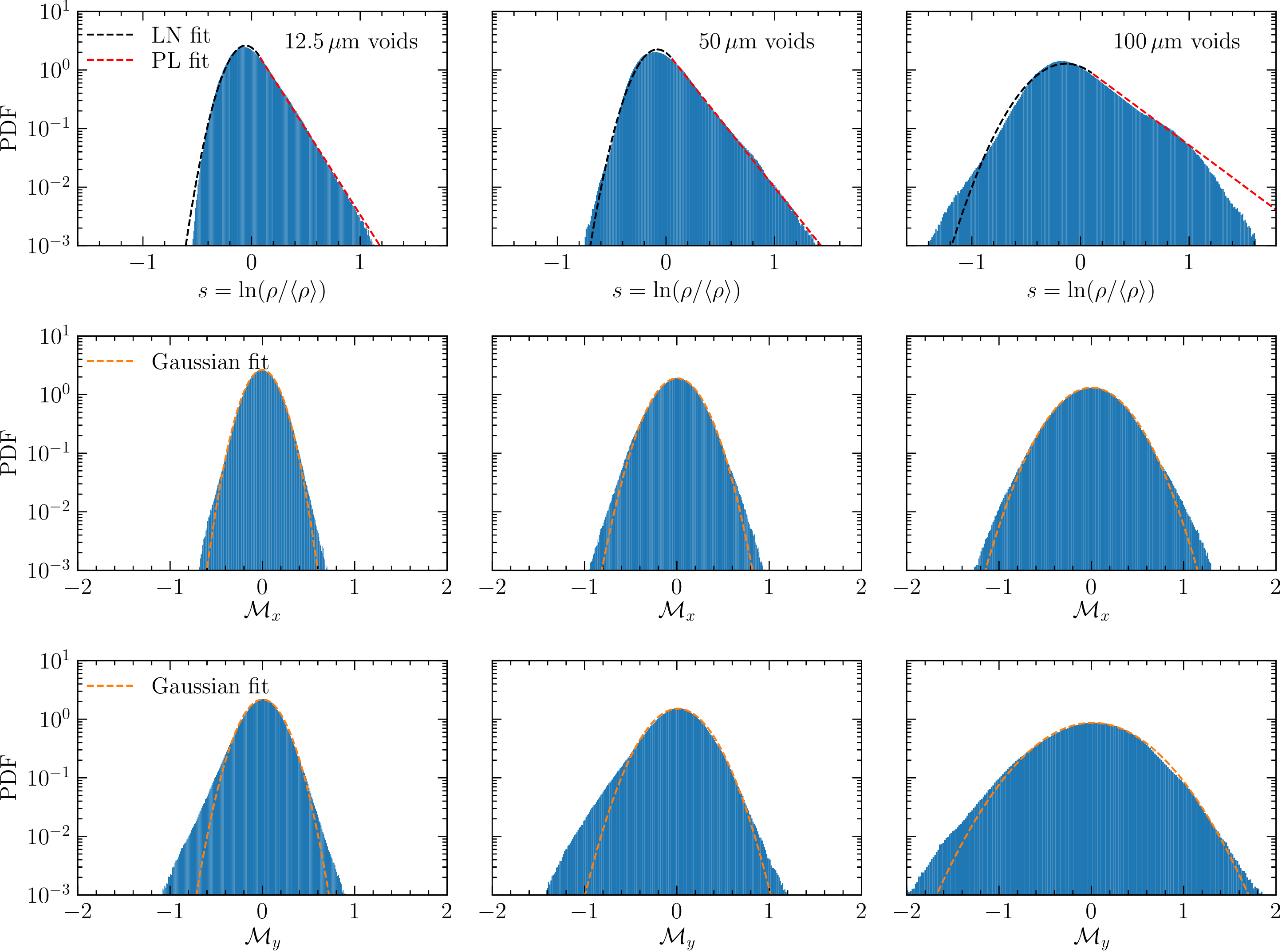}
    \caption{Comparison between the PDFs obtained for the three different foam void sizes: $12.5\,\mathrm{\mu m}$ (leftmost column), $50\,\mathrm{\mu m}$ (middle column) and $100\,\mathrm{\mu m}$ (rightmost column). The top panels show the PDFs of logarithmic density, $s$. The middle and bottom panels show the PDFs of the local Mach numbers along the $x$ and $y$ directions, respectively. The dashed curves represent LN+PL fits for the density PDFs and Gaussian fits for the Mach number PDFs.}
    \label{void_coparison}
\end{figure*}

\begin{figure*}
    \centering
    \includegraphics[width=\textwidth]{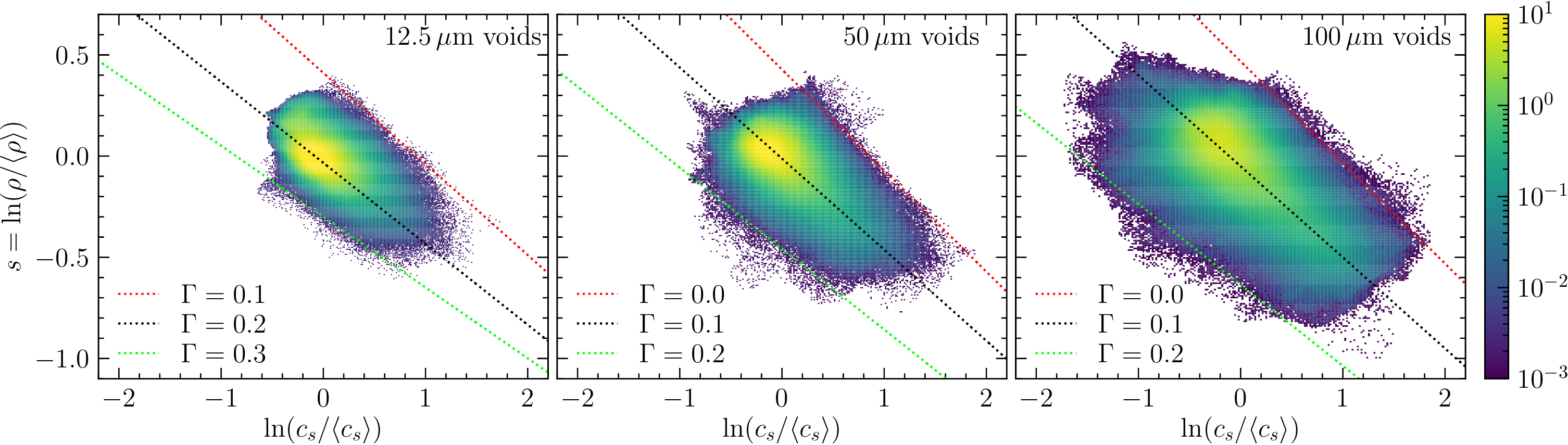}
    \caption{Density--sound speed correlation PDFs for the three different foam voids cases: $12.5\,\mathrm{\mu m}$, $50\,\mathrm{\mu m}$, and $100\mathrm{\mu m}$ (from left to right). Lines of constant polytropic $\Gamma$ are plotted to guide the eye. Three lines are drawn on each plot, one each at the top and the bottom edge of the PDF, and one passing approximately through the centre. We find that $\Gamma$ does not change significantly with different foam void size.}
    \label{rho_cs_all}
\end{figure*}

\begin{table*}
    \centering
    \caption{Comparison of density dispersion, Mach number, polytropic $\Gamma$, and driving parameter, in the three different foam void size cases.}
    \def\arraystretch{1.4}
    \setlength{\tabcolsep}{0pt}
    \begin{tabular*}{\linewidth}{@{\extracolsep{\fill}}ccccccccccc}
    \hline
    Simulation & Void size ($\mu$m) & $\sigma_{\rho/ \langle \rho \rangle }$& $\sigma_{\mathcal{M}_x}$ & $\sigma_{\mathcal{M}_y}$ & $\sigma_{\mathcal{M}_z}$ & $\mathcal{M}$ & $\Gamma$& $b$ \\
    (1) & (2) & (3) & (4) & (5) & (6) & (7) & (8) & (9)\\
    \hline
        Small & 12.5 & $0.21\pm 0.02$ & $0.15\pm 0.01$& $0.20\pm 0.01$ & $0.15\pm 0.01$ & $0.29\pm 0.01$ & 0.2$\pm$0.1 & $1.1 \pm 0.1$ \\
        Medium & 50 & $0.27\pm 0.01$ & $0.22\pm 0.01$ & $0.29\pm 0.01$ & $0.22\pm 0.01$ & $0.43\pm 0.01$ & 0.1$\pm$0.1 & $0.99\pm 0.07$ \\
        Large & 100 & $0.43\pm 0.03$ & $0.31\pm 0.02$ & $0.48\pm 0.05$ & $0.29\pm 0.01$ & $0.64\pm 0.04$ & 0.1$\pm$0.1& $0.93\pm 0.08$\\
    \hline
    \end{tabular*}
    \label{tab_void_comparison}
    {\raggedright \emph{Notes.} Columns: (1) simulation name, (2) diameter of foam voids, (3) standard deviation of density fluctuations, (4)--(6) standard deviation of the turbulent Mach number in the $x$, $y$, $z$ direction, (7) 3D turbulent Mach number computed via Eq.~(\ref{eq:3dmach}), (8) polytropic index $\Gamma$, (9) turbulence driving parameter (Eq.~\ref{eq:b_poly}). The errors in $\sigma_{\rho / \langle \rho \rangle}$, $\sigma_{\mathcal{M}_{x,y,z}}$, and $\mathcal{M}$ are calculated from the relative errors obtained by changing the analysis volume (see Tab.~\ref{tab:volume_dependence}). These errors are propagated analytically to obtain the uncertainties in $b$.
    \par}
\end{table*}

In order to study how the turbulence properties change with different foam void size, we now compare the previous $50\,\mu\mathrm{m}$ void size case with simulations using smaller ($12.5\,\mu\mathrm{m}$) and larger ($100\,\mu\mathrm{m}$) voids. Except for the void size, all other simulation parameters are kept unchanged. As in the case of $50\,\mu\mathrm{m}$ foam voids, we use the snapshot at $t=75\,\mathrm{ns}$ for the turbulence analysis. Appropriate positions for the analysis region are selected for each case, such that there is a sufficiently large volume of well-developed turbulence. The dimensions of the selected analysis region are the same for all three cases. The time evolution of all three simulations and the selected analysis regions are shown in Fig.~\ref{fig:void_comparison_snapshots}.

Fig.~\ref{void_coparison} shows the density PDFs (top panels) and Mach number PDFs (middle and bottom panels) for the three cases with different foam void sizes. Since the $\mathcal{M}_x$ and $\mathcal{M}_z$ PDFs are almost identical, we omit the $\mathcal{M}_z$ PDF in the figure and only show the $\mathcal{M}_x$ and $\mathcal{M}_y$ PDFs. We apply the LN+PL fit (Eq.~\ref{LNPL_eq}) to the density distributions, and the Gaussian fit (Eq.~\ref{mach_gaussian}) to the Mach number PDFs. We find that both the density and the Mach number PDFs become broader with increasing void size. This is because for larger void sizes, there is more space for the density and velocity fluctuations to develop when the shock passes through it. Relations between the post-shock properties and void size have been derived in a companion paper \citep{DavidovitsEtAl2022}.

The density PDFs in all the three cases are consistent with log-normal distributions at lower densities, with a high-density power-law tail that becomes flatter with increasing void size. When the size of the voids is small, the mixing of the two phases (foam and void material) is more efficient. Because of this, the density PDF for the smallest void size case is closest to a LN distribution, with a very steep power-law tail. With increase in the foam void size, the mixing becomes less and less efficient, there are more anisotropies, and the power-law tail flattens. The Mach number PDFs in all the three cases are close to Gaussian distributions, with non-Gaussian features in the form of wings far away from the mean, as discussed in Sec.~\ref{sec: mach}.

The parameters for the LN+PL fit of the density PDFs and for the Gaussian fits of the Mach number PDFs are summarised for all the three void sizes in Table~\ref{table:fit}. It is observed that the slope of the high-density power-law tail decreases with increasing foam void size, leading to a more prominent PL contribution. The standard deviation of the LN part ($\sigma_s$) increases with increasing void size, giving a more spread out PDF. The 3D Mach number also increases with increasing void size, but remains subsonic in all three cases.

Fig.~\ref{rho_cs_all} shows the density-sound speed correlation PDFs for the three cases. As expected from the density PDFs, the spread of the correlation PDFs also increases with increasing foam void size. All the three PDFs exhibit a large tail corresponding to regions with high densities and low sound speeds (hence low temperatures). There is clearly a negative correlation between the density and the sound speed for all the three cases. To determine the approximate polytropic $\Gamma$ of the systems, we plot lines of constant $\Gamma$ similar to the ones plotted before for the $50\,\mathrm{\mu m}$ void size case. We find that $\Gamma$ does not vary much with foam void size, with $\Gamma=0.2\pm 0.1$ for the $12.5\,\mathrm{\mu m}$ void size case, and $\Gamma=0.1\pm 0.1$ for the $50\,\mathrm{\mu m}$ and $100\,\mathrm{\mu m}$ void size cases, as discussed before in Sec.~\ref{sec: rho_cs_PDF}. Thus, the system has an effectively soft equation of state with $\Gamma < 1$, irrespective of the foam void size.

We calculate the density variance and 3D Mach number directly from the data for each void-size case. The values are listed in Table~\ref{tab_void_comparison}. We can also compare the values obtained directly from the data, with those obtained from the fits (see Tab.~\ref{table:fit}), and find that both $\sigma_{\rho/\langle\rho\rangle}$ and $\mathcal{M}$ agree between data and fits to within 10\% for all three void size cases. Using these measurements, we can now move on to the main objective of this study, which is to calculate the driving parameter of shock-driven turbulence.

\section{The driving parameter of shock-induced turbulence}
\label{sec: driving parameter} 

\begin{figure*}
	\includegraphics[width=\linewidth]{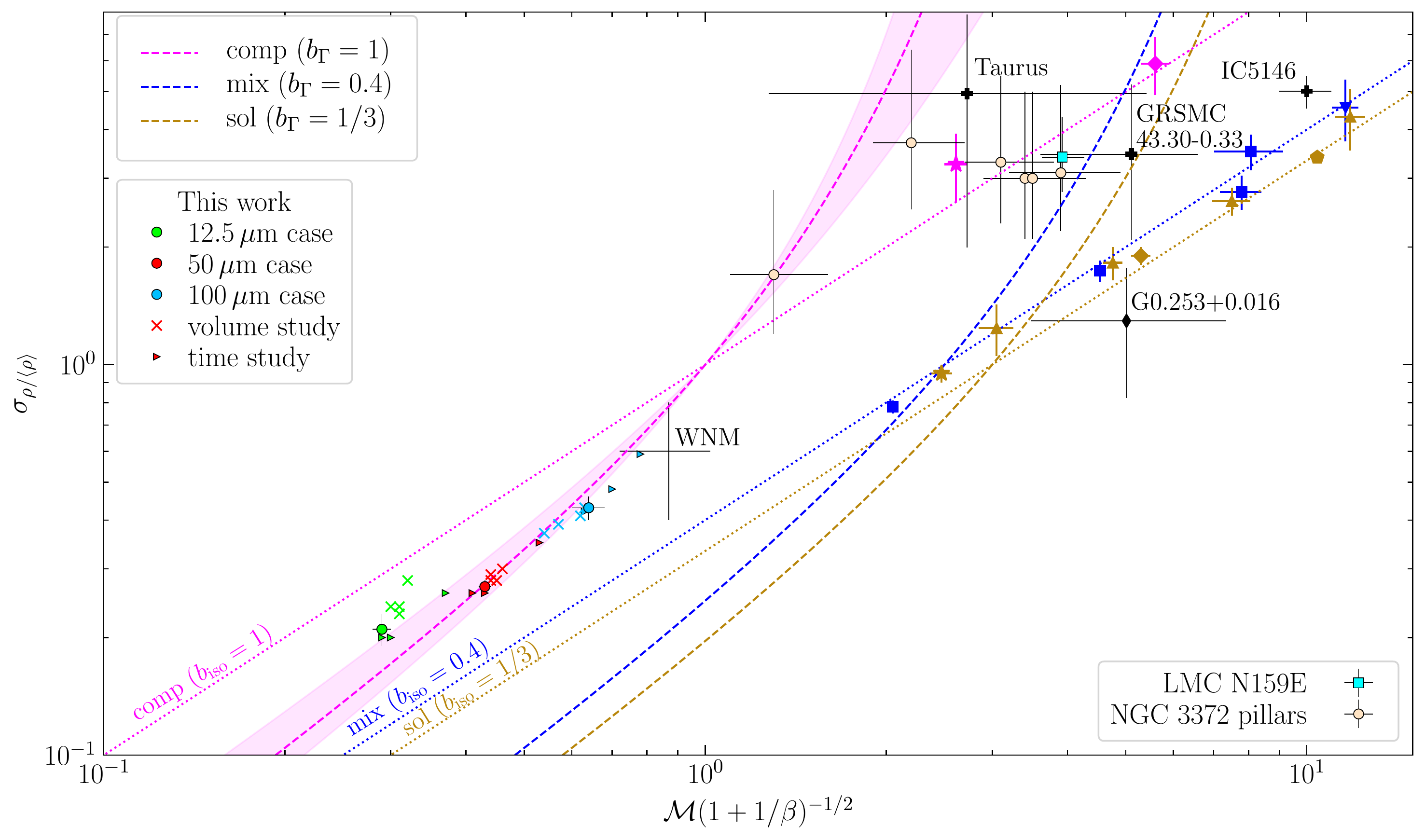}
    \caption{Relation between turbulent density fluctuations ($\sigma_{\rho/\langle\rho\rangle}$) and sonic Mach number ($\mathcal{M}$). For data points with magnetic field information (here only Taurus, G0.253+0.016, and the blue squares \citep{MolinaEtAl2012}), the sonic Mach number is modified by the plasma $\beta$ term from Eq.~(\ref{b_iso}), which simplifies to unity in the limit $\beta\to\infty$. The curves represent purely compressive driving ($b=1$, magenta), naturally mixed driving ($b \sim 0.4$, blue) and purely solenoidal driving ($b=1/3$, gold) for isothermal ($\Gamma=1$, dotted) and polytropic ($\Gamma=0.1$, dashed) cases, respectively, following Eq.~(\ref{eq:b_poly}). The magenta shaded area shows the region where $b_\Gamma=1$ for $0\leq \Gamma \leq 0.3$ (this is the range of $\Gamma$s in this study), again computed based on Eq.~(\ref{eq:b_poly}). The markers with black error-bars represent observational data points. These include three clouds in the spiral arms of the Milky Way: Taurus \citep{Brunt2010,KainulainenTan2013,FederrathEtAl2016}, GRSMC43.30-0.33 \citep{GinsburgFederrathDarling2013} and IC5146 \citep{PadoanJonesNordlund1997}, and the `Brick' cloud (G0.253+0.016) in the Central Molecular Zone \citep{FederrathEtAl2016}. The cream coloured circles show the six pillars analysed by \citet{MenonEtAl2021} in the Carina Nebula. The black cross labelled as `WNM' represents measurements in the Warm Neutral Medium (WNM) by \citet{MarchalEtAl2021}. The cyan square represents a measurement for the star-forming region N159E in the Large Magellanic Cloud \citep{ShardaEtAl2021b}. The gold, blue and magenta symbols are results of various numerical simulations, with the colour corresponding to the driving parameter of the turbulence used in the given simulation. These include results from studies by \citet{FederrathBanerjee2015} (inverted triangle), \citet{NolanFederrathSutherland2015} (triangles), \citet{KonstandinEtAl2012ApJ} (stars), \citet{MolinaEtAl2012} (squares), \citet{PriceFederrathBrunt2011,PriceFederrath2010} (pentagon) and \citet{FederrathKlessenSchmidt2008,FederrathDuvalKlessenSchmidtMacLow2010} (diamonds). The red circle shows the main result of our calculation for laser-induced, shock-driven turbulence for the $50\,\mathrm{\mu m}$ void foam. The $12.5\,\mathrm{\mu m}$ void size case is represented by the light green circle, whereas the $100\,\mathrm{\mu m}$ void size case is shown as the light blue circle. Red crosses correspond to the $50\,\mu$m void size case, but for different analysis volumes (see Appendix~\ref{sec: volume dependence}). The right-facing triangles show simulations obtained for each void size (using the corresponding colour), but for different analysis times (see Appendix~\ref{sec: time dependence}). Irrespective of the void size, analysis volume or time, we find that shock-driven turbulence is close to purely compressive driving ($b\sim1$).}
    \label{final_figure}
\end{figure*}

Many previous studies have shown that the standard deviation of the turbulent density fluctuations ($\sigma_{\rho/\langle \rho \rangle})$ and the 3D Mach number ($\mathcal{M})$ in a purely isothermal medium are related to each other through the relation \citep[e.g.,][]{PadoanNordlundJones1997,FederrathKlessenSchmidt2008,PriceFederrathBrunt2011,KonstandinEtAl2012ApJ,PadoanNordlund2011},
\begin{equation}
   \sigma_{\rho/\langle\rho\rangle}=b \mathcal{M} \left(1+1/\beta\right)^{-1/2}.
\label{b_iso}
\end{equation}
This is identical to Eq.~(\ref{eq:sigrho-mach}), but generalised to magnetised turbulence with plasma $\beta=p_\mathrm{th}/p_\mathrm{mag}$, i.e., the ratio of the thermal pressure to the magnetic pressure \citep{MolinaEtAl2012}. In the absence of magnetic fields, $p_\mathrm{mag}\to0 \implies \beta\to\infty$, simplifying Eq.~(\ref{b_iso}) to $\sigma_{\rho/\langle \rho \rangle}=b\mathcal{M}$, which is exactly Eq.~(\ref{eq:sigrho-mach}). In the following, we will ignore magnetic fields.

In Eqs.~(\ref{eq:sigrho-mach}) and~(\ref{b_iso}), the parameter $b$ is known as the turbulence driving parameter, which depends on the modes induced by the turbulent driving mechanism. As discussed in the introduction, some physical drivers of turbulence excite more solenoidal modes, other drivers excite more compressive modes \citep{Federrath2018}. The driving parameter covers the range \mbox{$b\sim1/3$--$1$}, with $b\sim1/3$ corresponding to a purely solenoidal (divergence-free) driving, and $b\sim1$ corresponding to a purely compressive (curl-free) driving. If the turbulence driving modes are randomly distributed in all the three directions, then $b\sim0.4$, which is known as the `natural mixture' \citep{FederrathKlessenSchmidt2008,FederrathDuvalKlessenSchmidtMacLow2010}. Thus, $b \lesssim 0.4$ means that there are more solenoidal modes as compared to compressive modes, whereas $b\gtrsim 0.4$ corresponds to more compressive modes as compared to solenoidal modes.

More recent studies have shown that Eq.~(\ref{b_iso}) is not applicable in non-isothermal conditions, and has to be modified \citep{NolanFederrathSutherland2015,FederrathBanerjee2015}. In particular, \citet{FederrathBanerjee2015} investigated the density variance -- Mach number relation in non-isothermal, polytropic turbulence ($p_\mathrm{th}\propto\rho^\Gamma$). In the absence of magnetic fields, the $\Gamma$-dependent relation for $b$ is given by \citep[combining eqs.~8 and~19 in][]{FederrathBanerjee2015},
\begin{equation}
    b_\Gamma = \left[\frac{\sigma_{\rho/\langle\rho\rangle}^{2\Gamma}-1}{\mathcal{M}^2\,\Gamma\left(1-\sigma_{\rho/\langle\rho\rangle}^{-2}\right)}\right]^{1/2}.
\label{eq:b_poly}
\end{equation}
We can immediately verify that for isothermal turbulence ($\Gamma=1$), Eq.~(\ref{eq:b_poly}) reduces to Eq.~(\ref{eq:sigrho-mach}). Thus, Eq.~(\ref{eq:b_poly}) is a generalised form for $b$ in gases with polytropic index $\Gamma$.

As quantified in Sec.~\ref{sec: rho_cs_PDF} and \ref{sec: void comparison}, the shock-driven turbulence investigated here has a range of polytropic $\Gamma$, the values of which vary only slightly with changes in the foam void size. Substituting the previously measured values of $\sigma_{\rho/\langle\rho\rangle}$, $\mathcal{M}$ and $\Gamma$, summarised in Table~\ref{tab_void_comparison}, into Eq.~(\ref{eq:b_poly}), we find $b=1.13\pm 0.08$, $0.99\pm 0.06$, and $0.93\pm 0.04$, for the $12.5\,\mathrm{\mu m}$, $50\,\mathrm{\mu m}$, and $100\,\mathrm{\mu m}$ void cases, respectively (listed in the last column of Table~\ref{tab_void_comparison}). Thus, the turbulence driven in all these cases is highly compressive, which is what one might expect for shock-driven turbulence. We also find that $b$ is nearly independent of the size of the structures in the medium, which is desirable, as $b$ is supposed to be a measure of the turbulence driving mechanism (i.e., here the laser-driven shock), and therefore, $b$ should be independent of the medium that is driven. However, there is a weak tendency for $b$ to increase with decreasing foam void size, although $b$ overlaps on a 1-sigma level between small and medium, and medium and large, respectively.

Close inspection of Eq.~(\ref{eq:b_poly}) shows that the value of $b_\Gamma$ is relatively mildly dependent on $\Gamma$. We recall that the values of the effective polytropic $\Gamma$ are in the range \mbox{$\Gamma=0.0$--$0.3$} (see Table~\ref{tab_void_comparison}), corresponding to a very soft effective equation of state. Thus, one might ask whether our results for $b$ are strongly dependent on $\Gamma$. To investigate this, we can make a rather extreme assumption and pretend that the gas is effectively isothermal ($\Gamma=1$), and use the simpler Eq.~(\ref{eq:sigrho-mach}) to evaluate $b$. Making this simple (incorrect) assumption, we would find $b= 0.72$, $0.63$, $0.67$, for the $12.5\,\mathrm{\mu m}$, $50\,\mathrm{\mu m}$, and $100\,\mathrm{\mu m}$ void cases, respectively. Thus, the driving parameters would still be significantly larger than the natural mix ($b\sim0.4$), and would therefore still indicate strongly compressive driving of the turbulence. However, use of Eq.~(\ref{eq:sigrho-mach}) is strongly discouraged in media with a polytropic index $\Gamma\ne1$, and Eq.~(\ref{eq:b_poly}) should be used instead, which indeed results in $b\sim1$ for the shock-driven turbulence simulations analysed here.

\section{Comparison with turbulence driving in the interstellar medium}
\label{sec: final}

Our numerical experiments suggest that shock-driven turbulence, such as supernova explosions or expanding radiation fronts driven by massive stars, may be considered compressive drivers, akin to the type of laser-induced, shock-driven turbulence in the numerical experiments presented here. Thus, we now want to place our simulation results into the bigger context of astrophysical observations, and compare them with observations in different environments, and with numerical studies related to astrophysical turbulence.

Fig.~\ref{final_figure} shows the turbulent density fluctuations, $\sigma_{\rho/\langle\rho\rangle}$, as a function of sonic Mach number\footnote{Data points for which magnetic field information is available, use the plasma $\beta$ term from Eq.~(\ref{b_iso}) in addition to the sonic Mach number, i.e., $\mathcal{M}\to\mathcal{M}(1+1/\beta)^{-1/2}$. Conversely, if the magnetic field is zero, $\beta\to\infty$.}, in a wide range of turbulent environments, from observations in the Warm Neutral Medium (WNM) over cold molecular clouds in the Milky Way, to the Large Magellanic Cloud, as well as previous idealised simulations of isothermal turbulence with different, controlled turbulence driving. The markers with black error-bars correspond to results from various astrophysical observations, whereas the coloured markers are data points from several numerical simulations. To guide the eye, we show lines of constant $b$ for the isothermal case ($\Gamma=1$), appropriate for the isothermal simulations shown as gold, blue and magenta data points, as well as for the polytropic case with $\Gamma=0.1$, appropriate for our shock-driven simulations presented above. We also show a shaded region, where $b=1$ and $\Gamma=[0.0, 0.3$], which is the range of $\Gamma$ found in our study of shock-driven turbulence.

The data points of the three different void size cases are represented by circles of different colours, with the $12.5$, $50$, and $100\,\mathrm{\mu m}$ void cases shown in green, red, and blue, respectively. The red crosses are the data points for the $50\,\mathrm{\mu m}$ void size case, but for different choices of the analysis volume (see Appendix~\ref{sec: time dependence}). The triangles, corresponding to each void size case (shown using the corresponding colour), represent the data points for different analysis times (see Appendix~\ref{sec: volume dependence}). As expected from the discussion in Sec.~\ref{sec: driving parameter}, all our data points lie very close to or inside the region with $b_\Gamma=1$, indicating strongly compressive driving, irrespective of the size of the voids in the foam or the choice of analysis region or time.

The astrophysical data points have a wide range of Mach numbers and driving parameters, depending on the different driving mechanisms and physical conditions. In molecular clouds, the gas is supersonic with Mach numbers \mbox{$\mathcal{M} \gtrsim 1$--$10$} \citep{SchneiderEtAl2013}. In our simulations of laser-driven turbulence, the Mach numbers are small and subsonic, because the terrestrial experiment can only achieve relatively small Mach numbers. Our study is much closer to the conditions in the Warm Neutral Medium (WNM) \citep{MarchalEtAl2021}, where the Mach numbers are of the order of unity or slightly subsonic. Depending on the appropriate value of $\Gamma$ for the WNM, $b$ is always greater than 0.5, suggesting that the turbulence driving in the WNM is predominantly compressive in nature.

It is worth noting that although some of the physical parameters, such as the absolute temperature, velocity, size and time scales in our numerical and laboratory experiments are starkly different from those observed in the ISM, the dimensionless parameters of the turbulence -- that is, the 3D Mach number ($\mathcal{M}$), the relative density fluctuations ($\sigma_{\rho/\langle\rho\rangle}$), and the turbulence driving parameter ($b$) -- can be directly compared. In summary, we find that shocks are generally compressive drivers of turbulence ($b\sim1$). Since many drivers of turbulence in the ISM fall into the category of shock drivers (SN explosions, stellar winds and ionisation, gravitational collapse, spiral-arm compression, bow shocks from jets, etc.), we may expect compressive drivers of turbulence in many astrophysical environments in the ISM, with important implications for the structure and star formation activity of interstellar clouds. It should be noted, however, that our simulations are highly simplified, with voids of constant radius in each simulation set. The ISM has a much more complex density structure. However, our finding that the value of the driving parameter is relatively insensitive to the size of the structures in the medium, lends support to the robustness of the main result of this study, namely that the driving mode of shock-driven turbulence is compressive, with a $b$-parameter of $\sim1$.

\section{Summary and conclusions}
\label{sec: summary}
In this study, we performed a hydrodynamic simulation of laser-induced, shock-driven turbulence in a pre-structured multi-phase medium. We analysed regions of well-developed turbulence driven by the shock, determined the properties of the turbulence and calculated the turbulence driving parameter. The dependence of the turbulence properties on the size of the structures in the medium was also studied. We now summarise the main results of this study:
\begin{enumerate}
    \item The density PDF is consistent with a log-normal distribution at lower densities and exhibits a high-density power-law tail. The high-density power-law tail can be attributed to the multi-phase nature of the medium. The sound speed PDF shows a deformed Gaussian shape with an exponential tail at lower values of sound speed, corresponding to colder regions.
    
    \item There is a negative correlation between the sound speed and the density. This is a result of a combination of the initial pressure equilibrium and the multi-phase nature of the medium in the simulations. Initially, the foam, which is denser, is colder as compared to the less dense air in the voids, because of pressure equilibrium. After the shock passes, the two phases mix and the medium becomes turbulent. High-density regions stay cooler as compared to the low-density regions, giving rise to the negative correlation between density and sound speed. The overall hydrodynamics of the turbulence are similar to that of a system with polytropic \mbox{$\Gamma\sim0.0$--$0.3$}, with slight variation depending on the size of the voids in the foam.
    
    \item The local Mach number PDFs are consistent with a Gaussian shape, which is a characteristic of a purely turbulent medium. Because of the $xz$-symmetry of the setup, the $\mathcal{M}_x$ and $\mathcal{M}_z$ PDFs are almost identical, with the $\mathcal{M}_y$ PDF being broader by \mbox{$\sim30$--$60\%$}, due to the stronger velocity fluctuations induced along the shock propagation direction. The turbulence remains subsonic with a 3D Mach number \mbox{$\mathcal{M}\sim0.3$--$0.6$}, increasing with increasing foam void size.
    
    \item The turbulent density fluctuations ($\sigma_{\rho/\langle \rho \rangle}$) and the 3D Mach number were calculated directly from the data as well as from the fits applied to the PDFs. The values obtained from the two methods were found to be very similar, with deviations of $\lesssim 10\%$.
    
    \item We calculated the turbulence driving parameter using the density variance -- Mach number relation, Eq.~(\ref{eq:b_poly}), for a polytropic equation of state. The $b$ parameter for the foam with $50\,\mathrm{\mu m}$ voids is consistent with purely compressive driving, and is only mildly dependent on the value of the polytropic $\Gamma$.
    
    \item We studied the dependence of the turbulence properties on the size of the structures in the medium by considering three different simulations with different void sizes. It was observed that the overall shape of the density and Mach number PDFs remains unchanged with change in void size. However, the spread of the PDFs increases with increase in void size, and the power-law part of the density PDF becomes more noticeable due to poorer mixing when the void structures become larger. The increase in the standard deviation of the density and the Mach number PDFs is such that the turbulence driving parameter remains almost unchanged, with a weak tendency of $b$ decreasing with increasing foam void size; yet, there is 1-sigma overlap in $b$ between the small and medium, and medium and large voids cases, respectively.
    
    \item The value of the driving parameter is always much greater than $0.4$ (corresponding to a natural mixture) -- irrespective of the void size, choice of analysis volume or time -- indicating a strongly compressive driving induced by shocks.
    
    \item We compare our numerical simulations with other simulations, and with a wide range of observations in molecular clouds and in the Warm Neutral Medium. We conclude that shock-driven turbulence in a pre-structured multi-phase medium is consistent with highly compressive driving, which may have implications for the predominant structures and star formation mode in the ISM.
\end{enumerate}

\section*{Acknowledgements}
We thank Kumar Raman, Mario Manuel, and Nino Landen for supporting this project. We acknowledge the NIF Discovery Science Program for allocating upcoming facility time on the NIF Laser to test aspects of the models and simulations discussed in this paper. C.~F.~acknowledges funding provided by the Australian Research Council (Future Fellowship FT180100495), and the Australia-Germany Joint Research Cooperation Scheme (UA-DAAD). The work of S.~D.~was supported in part by the LLNL-LDRD Program under Project No.~20-ERD-058. Work by LLNL was performed under the auspices of U.S.~DOE under Award No.~DE-AC52-07NA27344. We further acknowledge high-performance computing resources provided by the Australian National Computational Infrastructure (grant~ek9) in the framework of the National Computational Merit Allocation Scheme and the ANU Merit Allocation Scheme, and by the Leibniz Rechenzentrum and the Gauss Centre for Supercomputing (grant~pr32lo). The simulation software FLASH was in part developed by the DOE-supported Flash Center for Computational Science at the University of Chicago.

\section*{Data Availability}
The simulation data in this article will be shared upon reasonable request to Christoph Federrath (\href{mailto:christoph.federrath@anu.edu.au}{christoph.federrath@anu.edu.au}).
 


\def\rmp{{Rev.~Mod.~Phys.}}
\def\jfm{{J.~Fluid~Mech.}}
\bibliographystyle{mnras}
\bibliography{federrath}



\appendix

\section{1D shock preparation} \label{sec:1dshock}

The 3D simulations were initialized with a density, velocity and pressure profile obtained with a 1D Lagrangian code, specifically developed to prepare laser experiments by running radiative hydrodynamics multi-material simulations \citep{BenuzziEtAl2001}. This CHARM code features a fully implicit Lagrangian solver with multi-group radiative transfer, different electron and ion temperature, laser energy deposition, thermal electronic conduction, ionic viscosity and SESAME equation of state for multiple materials. Opacities for each separate material are computed using a simple average atom model. This code has proven accurate enough to model laser experiments with direct and indirect drives on the Phebus laser \citep{BenuzziEtAl2001} and on the Omega laser \citep{KaneEtAl2001}.

\begin{figure}
    \centering
    \includegraphics[width=\linewidth]{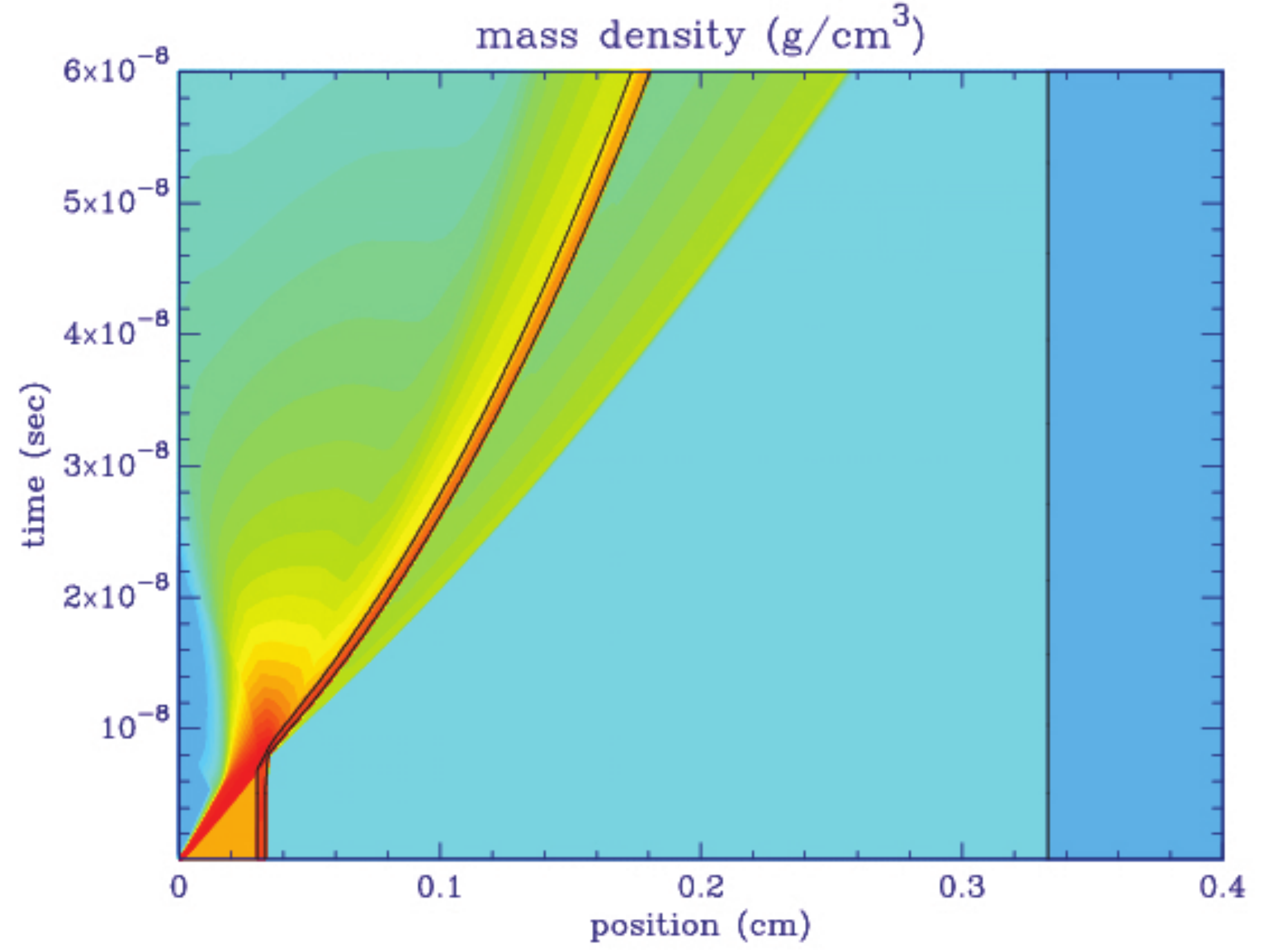}
    \caption{Space-time diagram of the material density. The colour map is logarithmically spaced between density of $0.01\,\mathrm{g/cm^3}$ and $10\,\mathrm{g/cm^3}$.}
    \label{fig:1dshock_spacetime}
\end{figure}

Figure~\ref{fig:1dshock_spacetime} shows the space-time material density diagram of the setup, where one can see the initial multi-layer plastic, Aluminum and foam target. The basic evolution is as follows. The laser direct drive launches a strong shock in the plastic ablator, which impacts the thin Aluminum layer at $t=6\,\mathrm{ns}$. The shock is then transmitted in the foam, while the Aluminum foil trajectory is slowly decelerated by the foam.

\begin{figure}
    \centering
    \includegraphics[width=\linewidth]{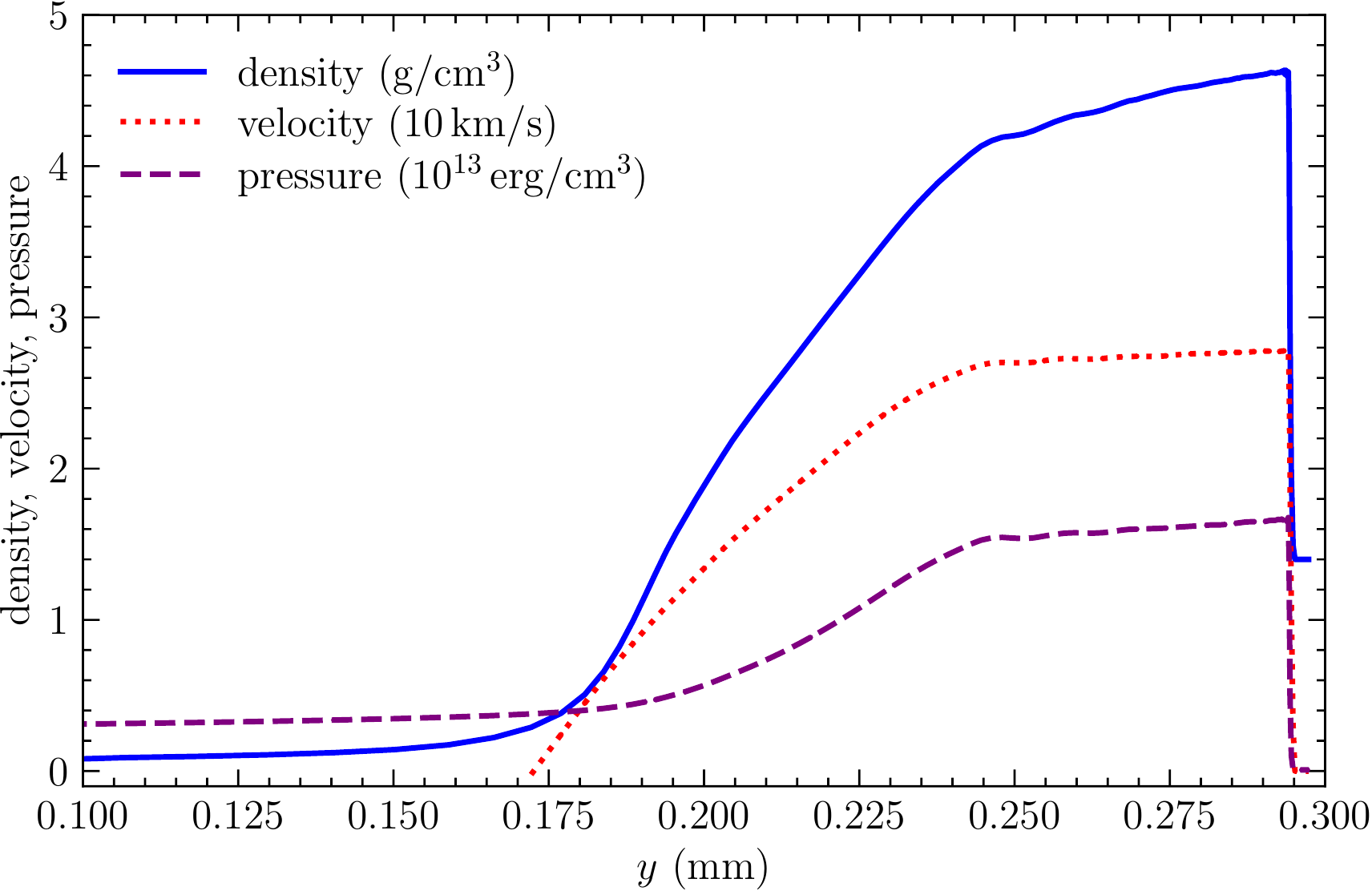}
    \caption{Density, velocity, and pressure profile of the 1D shock generated with the CHARM code, which we use to initialise the shock in the plastic ablator in our 3D simulations (c.f., Fig.~\ref{fig:initial_conditions}).}
    \label{fig:1dshock_profile}
\end{figure}

For the purposes of initialising the shock in the 3D simulations, we take the 1D shock profile from here. Figure~\ref{fig:1dshock_profile} shows the shock profile after the laser drive has been switched off, but before the shock has reached the Aluminum foil, which happens at $t=5\,\mathrm{ns}$ in Fig.~\ref{fig:1dshock_spacetime}. We use this density, velocity, and pressure profile to initialise the shock inside the plastic ablator for the 3D simulations presented in the main part of the study (see Fig.~\ref{fig:initial_conditions}).

\section{Time evolution of density structures}
\begin{figure*}
    \centering
    \includegraphics[width=\linewidth]{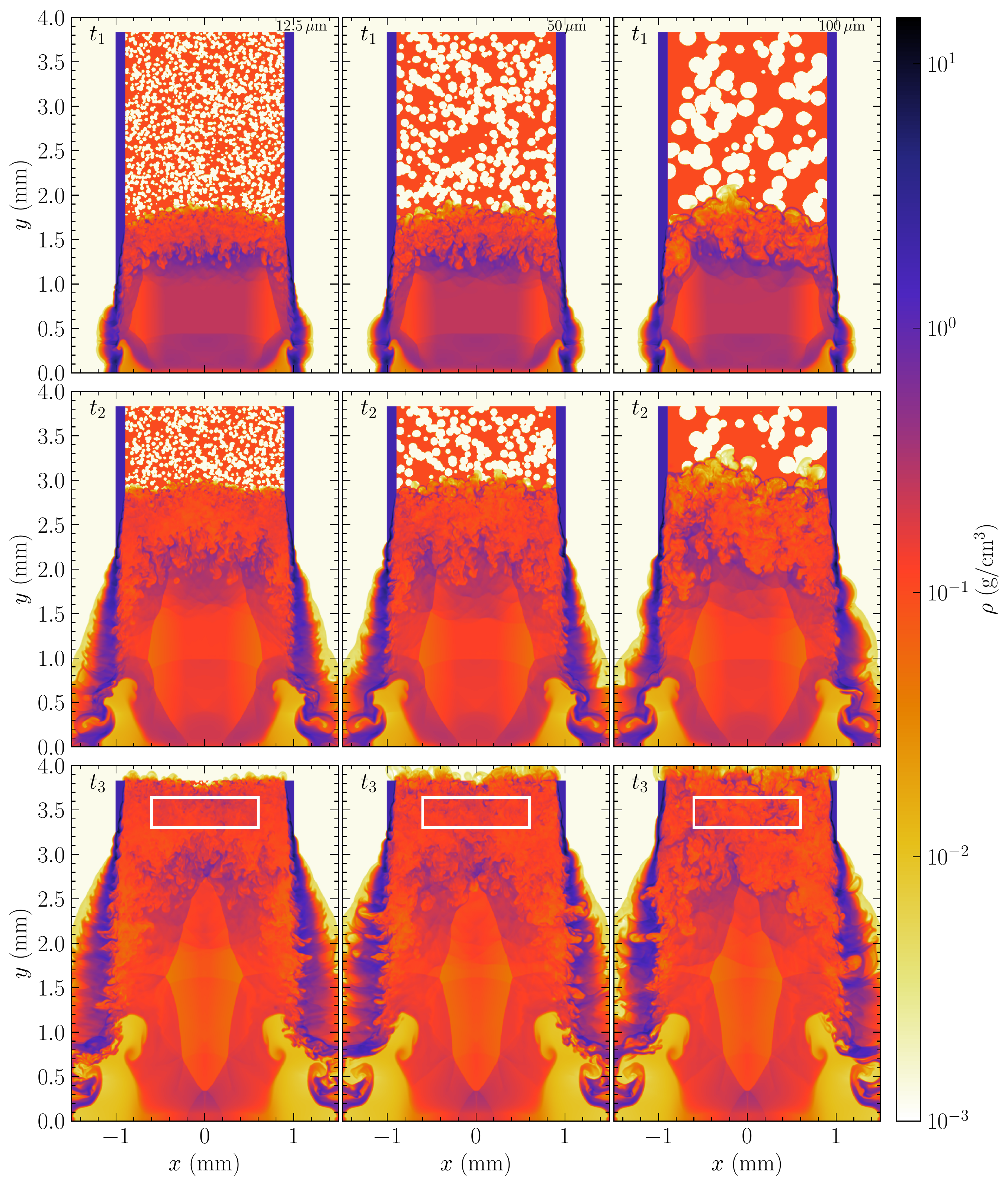}
    \caption{Same as the top panels of Fig.~\ref{fig:time_evolution}, but here we compare the time evolution of all three simulations with foam void sizes of $12.5\,\mathrm{\mu m}$, $50\,\mathrm{\mu m}$, and $100\,\mathrm{\mu m}$ (from left to right), at times $t_1=25\,\mathrm{ns}$ (top panels), $t_2=50\,\mathrm{ns}$ (middle panels) and $t_3=75\,\mathrm{ns}$ (bottom panel). The turbulence analysis is carried out at $t=75\,\mathrm{ns}$ in all three cases. For the analysis we selected a sufficiently large region of well-developed turbulence, indicated by the white rectangles.}
    \label{fig:void_comparison_snapshots}
\end{figure*}

Fig.~\ref{fig:void_comparison_snapshots} shows the time evolution of the density structures as the shock passes through the foam, for all three simulations with different foam void sizes. We show $xy$-slices of the density at three different time instances after the initiation of the shock. The bottom panels show the time at which the analysis is carried out ($t=75\,\mathrm{ns}$), with the white rectangles representing the position of the analysis regions.

\section{Sound speed PDF fit and summary of fit parameters}
\label{fit_appendix}
\begin{table*}
    \centering
    \caption{Fit parameters of the density PDF $p(s)$ (Eq.~\ref{LNPL_eq}), sound speed PDF $p(c_s/\langle c_s \rangle)$, and Mach number PDFs $p(\mathcal{M}_{x,y,z})$ (Eq.~\ref{mach_gaussian}), for the 3~foam void sizes.}
    \def\arraystretch{1.4}
    \begin{tabular*}{\linewidth}{@{\extracolsep{\fill}}ccccccccc}
    \hline
    Quantity& Symbol & $12.5\,\mathrm{\mu m}$ foam voids & $50\,\mathrm{\mu m}$ foam voids & $100\,\mathrm{\mu m}$ foam voids \\ 
    \hline
    Standard deviation of the LN part of $p(s)$ & $\sigma_s$ &$0.143^{+0.003}_{-0.004}$ & $0.159^{+0.007}_{-0.005}$ & $ 0.265^{+0.010}_{-0.007}$\\
    Transition point between the LN and the PL & $s_t$ & $0.080^{+0.013}_{- 0.014}$ & $0.054^{+0.018}_{-0.013}$& $0.075^{+ 0.030}_{- 0.021}$\\
    Slope of the PL part of $p(s)$ & $\alpha$ & $6.65^{+ 0.15}_{- 0.13}$ & $5.31^{+0.10}_{-0.08}$ & $3.13^{+ 0.06}_{- 0.05}$\\
    Standard deviation of the Gaussian part of $p(c_s/\langle c_s \rangle)$ & $\sigma_{c_s}$ & $0.066^{+0.001}_{- 0.002}$ & $0.076^{0.001}_{0.002}$ & $ 0.123^{+0.002}_{+0.002}$ \\
    Transition point between the Exponential and the Gaussian & $c_{s_t}$& $0.897^{+ 0.004}_{- 0.003}$ & $0.950^{+ 0.003}_{- 0.003}$ & $0.935^{+ 0.005}_{- 0.006}$\\
    Slope of the exponential part of $p(c_s/\langle c_s \rangle)$ & $\nu$ & $22.63^{+ 0.43}_{- 0.43}$ & $12.05^{+ 0.17}_{- 0.18}$ & $6.43^{+ 0.10}_{- 0.12}$\\
    Standard deviation of $p(\mathcal{M}_x)$ & $\sigma_{\mathcal{M}_x}$ & $0.151^{+0.002}_{-0.002}$ & $0.211^ {+0.003}_{-0.003}$ & $0.304^{+0.004}_{-0.004} $\\
    Mean of $p(\mathcal{M}_x)$ & $\langle \mathcal{M}_x \rangle$ & $-0.002^{+0.003}_{-0.003}$ & $0.003^{+0.004}_{-0.004}$ & $0.001^{+0.006}_{-0.005}$\\
    Standard deviation of $p(\mathcal{M}_y)$ & $\sigma_{\mathcal{M}_y}$& $0.184^{+0.003}_{-0.003}$ & $0.263^{+ 0.004}_{-0.004}$ & $0.459^{+0.007}_{-0.006}$\\
    Mean of $p(\mathcal{M}_y)$ & $\langle \mathcal{M}_y \rangle$ & $0.003^{+0.003}_{-0.003}$& $ 0.011^{+0.005}_{-0.004}$ & $ 0.014^{+0.008}_{-0.008}$\\
    Standard deviation of $p(\mathcal{M}_z)$ & $\sigma_{\mathcal{M}_z}$ & $0.150^{+0.002}_{-0.002} $ & $0.205^{+0.003}_{-0.003}$ & $0.291^{+0.004}_{-0.004}$\\
    Mean of $p(\mathcal{M}_z)$ & $\langle \mathcal{M}_z \rangle$ & $0.000^{+0.003}_{-0.003}$& $0.003^{+0.004}_{-0.004}$ & $-0.005^{+0.005}_{-0.005}$\\
    \hline
    Total density fluctuations based on Eq.~(\ref{eq:pdfintegral}) & $\sigma_{\rho/\langle\rho\rangle}$ & $0.211^{+ 0.002}_{- 0.002}$ & $0.269^{+ 0.004}_{- 0.003}$ & $0.456^{+0.004}_{- 0.005}$\\
    3D Mach number (Eq.~\ref{eq:3dmach}) & $\mathcal{M}$& $0.281^{+0.003}_{-0.002}$ & $0.394^{+0.004}_{-0.003}$ & $0.623^{+0.005}_{-0.006}$\\
    Turbulence driving parameter (Eq.~\ref{eq:b_poly}) & $b_\Gamma$ & $1.17^{+0.08}_{-0.08}$ & $1.08^{+0.07}_{-0.07}$ & $0.99^{+0.04}_{-0.04}$\\
    \hline
    \end{tabular*}
    \label{table:fit}
    {\raggedright \emph{Notes.} The last three rows are derived quantities from the fitted parameters. For $\sigma_{\rho/\langle\rho\rangle}$ we integrate the numerical PDF via Eq.~(\ref{eq:pdfintegral}) up to $s=1.5$. This upper bound is only relevant for the $100\,\mathrm{\mu m}$ case, because the PL section is very flat there, and integrating to infinity would yield unreasonably large $\sigma_{\rho/\langle\rho\rangle}$.
    \par}
\end{table*}

Analogous to the LN+PL fit used for the density PDF, the sound speed PDF is fitted using an Exponential + Gaussian fit:
\begin{equation}
\label{cs_fit}
    p(c_s/\langle c_s \rangle)=
    \begin{cases}
    A\exp{(\nu c_s/\langle c_s \rangle)} \hspace{1cm} &\text{ for } s\ge c_{s_t}\\
    \frac{B}{\sqrt{2 \pi \sigma_{c_s}^2}}\exp{\frac{(c_s/ \langle c_s \rangle-c_{s0})^2}{2\sigma_{c_s}^2 }} \hspace{1cm} &\text{ for }c_s\leq c_{s_t}
    \end{cases}
\end{equation}
Three constraints are applied to the fit: the normalisation condition and the continuity of the function ($p(c_s/\langle c_s \rangle$) and its derivative ($dp(c_s/\langle c_s \rangle)/d(c_s/\langle c_s \rangle)$) at the point of transition from the exponential to the Gaussian part, $c_{s_g}$. Because of the constraints, there are only three free parameters, which are chosen to be the slope of the exponential part in the logarithmic PDF ($\nu$), the width of the Gaussian ($\sigma_{c_s}$) and the transition point ($c_{s_t}$). The mean of the Gaussian ($c_s0$) and the normalisation constants, $A$ and $B$ can be derived from these parameters using the constraint conditions. As seen from Fig.~\ref{cs_PDF}, the exponential + Gaussian is a reasonably good fit for the sound speed PDF, with deviations only at very low sound speeds.\\

Table~\ref{table:fit} presents the parameters of the density, sound speed and Mach number fits for all the three cases considered. Monte Carlo sampling was used to determine the errors in the fit parameters, where $\sim500$ PDFs were fitted, each with $\sim10000$ randomly sampled points from the analysis region. It is observed that the slopes of the over-density tail and the low sound speed tail decrease with increase in the void size, making the power law and exponential more dominant. The width of the LN part in the density PDF and Gaussian in the sound speed PDF increase with increase in the void size, though the difference between the $12.5 \, \mathrm{\mu m}$ and $50 \, \mathrm{\mu m}$ cases is not much. The values of $\mathcal{M}_x$ and $\mathcal{M}_z$ are almost identical for each of the three cases, which is a result of the $xz$ symmetry. The 3D Mach numbers obtained from the data and from the Gaussian fits are in good agreement. Thus, non-Gaussian features of the Mach number PDFs can be neglected.

\section{Velocity and density gradients}
\label{sec: gradients}

\begin{figure*}
    \centering
    \includegraphics[width=\linewidth]{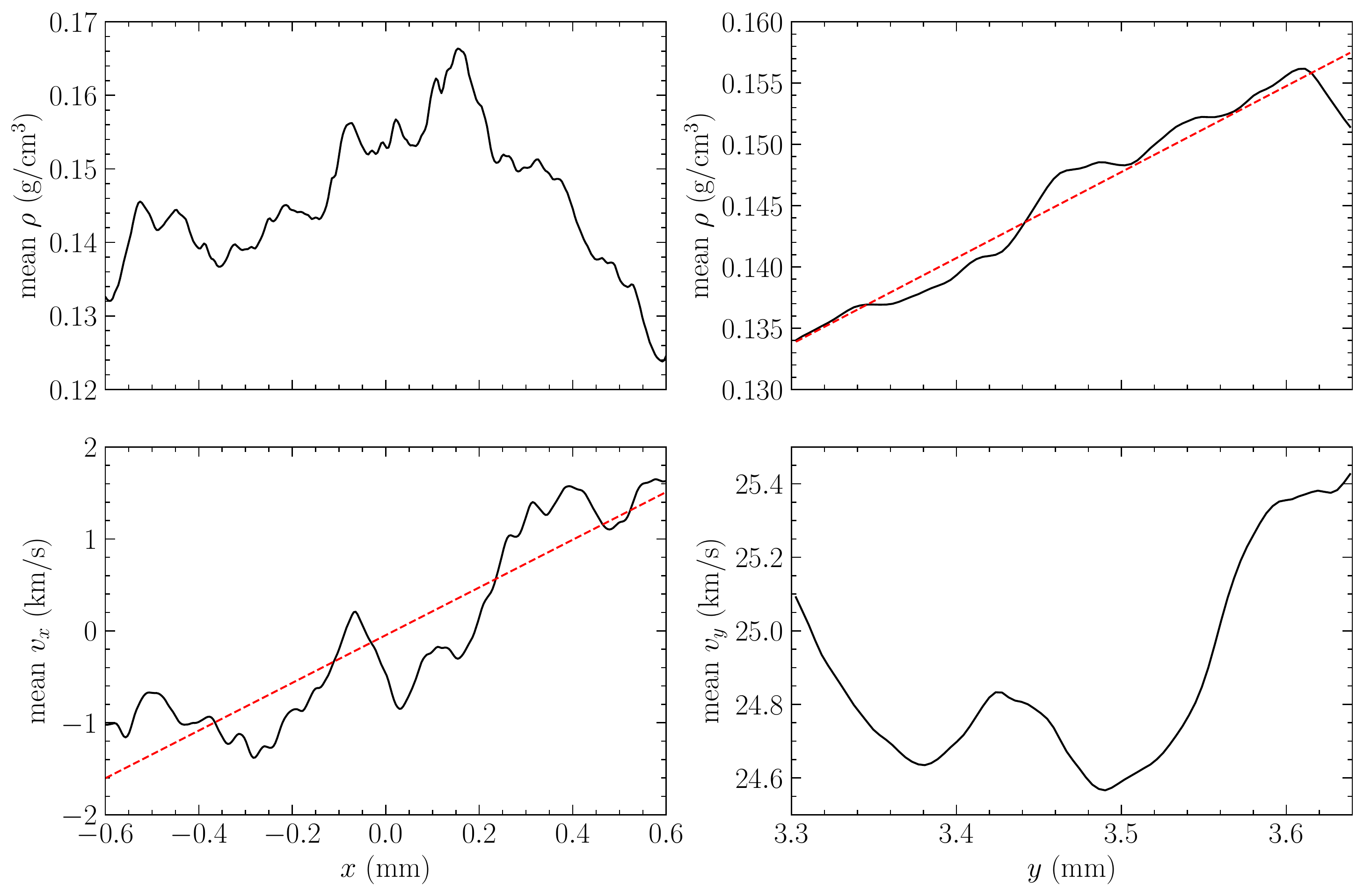}
    \caption{Plots showing the gradients in the density (top panels) and the velocity (bottom panels) in the analysis region along $x$-direction (left panels) and $y$-direction (right panels) respectively. It is seen that there is a net density gradient along the direction of the shock propagation, and a gradient in the $x$-component of the velocity along the $x$-direction. These two gradients are fitted with a linear fit, represented by red dashed lines. Overall, the contribution of the gradients is negligible as compared to the turbulent density and velocity fluctuations.}
    \label{fig: gradients}
\end{figure*}

As the shock moves through the foam, the foam and the walls are pushed outwards. Thus, the particles towards left move leftwards and the particles towards right move rightwards. Because of this, there are gradients in the x and z components of velocity ($v_x$ and $v_z$) along $x$ and $y$ directions respectively.\\
In order to look to at the velocity gradients and to correct for them, we take $yz$ slices of density for each value of $x$-position in the analysis volume. We calculate the mean $v_x$ for each slice, and plot it as a function of position. The bottom left panel in Fig.~\ref{fig: gradients} shows the $v_x$ gradient which is fitted using a linear fit. It is observed that the gradient is negligible as compared to the overall turbulent $v_x$ fluctuations in the analysis region. We subtracted the $v_x$ gradient from the $v_x$ data, and it was found that the Mach-number PDF almost remains unchanged. Similar results were obtained for the $v_z$ gradient along $z$-direction. Thus, the gradients in x and z components of the velocity have negligible effect on the turbulence properties and are neglected in the final analysis.\\

Since the shock propagates along the $y$-direction, it is observed that there is a gradient in the density along this direction. We use a procedure similar to the one used for velocity gradients for studying the density gradient. The top right panel in Fig.~\ref{fig: gradients} shows the gradient in density along $y$-direction, which is fitted using a linear fit. The density gradient is negligible as compared to the overall turbulent density fluctuations, and correcting for it keeps the density PDF almost unchanged. Therefore, the density gradient is also neglected in the final analysis.

\section{Volume dependence of the turbulence properties}
\label{sec: volume dependence}

\begin{table*}
    \centering
    \caption{Volume dependence of the turbulence properties}
    \def\arraystretch{1.3}
    \setlength{\tabcolsep}{0pt}
    \begin{tabular*}{\linewidth}{@{\extracolsep{\fill}}ccccccccc}
    \hline
    Void size ($\mu$m)& Relative volume & $\sigma_{\rho/ \langle \rho \rangle }$& $\sigma_{\mathcal{M}_x}$ & $\sigma_{\mathcal{M}_y}$ & $\sigma_{\mathcal{M}_z}$ & $\mathcal{M}$ & $b$ \\
    \hline
        12.5 &1& 0.21	&0.15&	0.20	&0.15& 0.29 &$1.13^{+0.08}_{-0.08}$\\
        & 0.5 (x is halved)  & 0.24	&0.15 & 0.21 &0.16 & 0.30 & $1.22^{+0.08}_{-0.08}$ \\
        & 0.5( z is halved) &0.24 & 0.16 & 0.22 & 0.15 & 0.31& $1.18^{+0.08}_{-0.08}$ \\
        & 0.25 (both x and z are halved) & 0.28	& 0.15 & 0.23 & 0.16&	0.32  &$1.29^{+0.08}_{-0.07}$\\
        & 0.125 (all sides are halved) & 0.23 & 0.15 & 0.22 & 0.16&	0.31& $1.14^{+0.08}_{-0.07}$\\
    \hline
        50 & 1 & 0.27 & 0.22 & 0.29 & 0.22 & 0.43 & $0.99^{+0.06}_{-0.06}$\\
        & 0.5 (x is halved)  & 0.28 & 0.22 & 0.30 & 0.23 & 0.44 & $0.99^{+0.07}_{-0.05}$ \\
        & 0.5( z is halved) & 0.29 & 0.22 & 0.31 & 0.22 & 0.44 & $1.02^{+0.06}_{-0.06}$ \\
        & 0.25 (both x and z are halved) & 0.30 & 0.23 & 0.33 & 0.23 & 0.46 &$1.00^{+0.06}_{-0.06}$\\
        & 0.125 (all sides are halved) & 0.28 & 0.22 & 0.33 & 0.22 & 0.45 & $0.97^{+0.06}_{-0.06}$\\
    \hline
        100 & 1 & 0.43 & 0.31 & 0.48 & 0.29 & 0.64& $0.93^{+0.04}_{-0.04}$\\
        & 0.5 (x is halved)  & 0.43 & 0.30 & 0.46 & 0.30 & 0.63& $0.94^{+0.04}_{-0.04}$ \\
        & 0.5( z is halved) & 0.41 & 0.30 & 0.46 & 0.29 & 0.62& $0.93^{+0.04}_{-0.04}$ \\
        & 0.25 (both x and z are halved) & 0.39 & 0.3 & 0.4 & 0.28 & 0.57 &$0.97^{+0.05}_{-0.04}$\\
        & 0.125 (all sides are halved) & 0.37 & 0.27 & 0.37 & 0.28 & 0.54 & $0.99^{+0.05}_{-0.05}$\\
    \hline
    \end{tabular*}
    \label{tab:volume_dependence}
\end{table*}

Here, we investigate whether the properties of the turbulence depend on the choice of volume for the analysis region considered. We take different analysis volumes centred on the same point ($x=0.0\,$mm, $y=3.47\,$mm, $z=0.0\,$mm) for all the three void size cases. Five different cases are considered for each void size. Starting from the region for the main analysis, we first halve one side of the cuboidal region, then halve two sides, and finally halve all the sides, so that the final analysis region has one eighth of the original volume. Table~\ref{tab:volume_dependence} shows the results obtained. The polytropic $\Gamma$ for the system does not change with changes in the analysis volume. It is seen that all the values of interest change only slightly with variations in the analysis volume. In fact, the variation in the driving parameter is less than $15\%$ for all the three void size cases considered. Thus, we conclude that the turbulence properties measured remain largely independent of the choice of the analysis volume as long as the volume is not too small (due to statistical limitations) or too large (when the density and velocity gradients and/or boundary effects might become important). 

\section{Time dependence of the turbulence properties}
\label{sec: time dependence}
\begin{table*}
    \centering
    \caption{Time dependence of the turbulence properties.}
    \def\arraystretch{1.3}
    \setlength{\tabcolsep}{0pt}
    \begin{tabular*}{\linewidth}{@{\extracolsep{\fill}}cccccccccc}
    \hline
    Void size ($\mu$m)&Time (ns)  & $y$ (mm) & $\sigma_{\rho/ \langle \rho \rangle }$& $\sigma_{\mathcal{M}_x}$ & $\sigma_{\mathcal{M}_y}$ & $\sigma_{\mathcal{M}_z}$ & $\mathcal{M}$ & $\Gamma$ & $b$  \\
    \hline
        12.5 & 50 & 2.6 & 0.26 & 0.19 & 0.26 &0.19 & 0.37 &  $0.0\pm 0.1$ & $1.19^{+0.10}_{-0.08}$\\
         & 65 & 3.1 & 0.20 & 0.15 & 0.21 & 0.15 & 0.30 & $0.2\pm0.1$ & $1.05^{+0.08}_{-0.07}$\\
         & 75 & 3.47 & 0.20 & 0.15 & 0.20 &0.15 & 0.29 &  $0.2 \pm 0.1$ &$1.08^{+0.08}_{-0.07}$ \\
        \hline
        50&50 & 2.6 & 0.35 & 0.26 & 0.38 & 0.27 & 0.53 & $0.1\pm 0.1$ & $0.97^{+0.05}_{-0.05}$\\
        &65& 3.1 & 0.26 & 0.22 & 0.30 & 0.22 & 0.43 &  $0.1\pm 0.1$ &$0.96_{-0.06}^{+0.06}$\\
         &75 & 3.47 & 0.26& 0.22 & 0.28 & 0.21 & 0.41 & $0.1\pm0.1$& $1.01^{+0.07}_{-0.06}$\\
         \hline
         100& 50 & 2.6 & 0.59 & 0.38 & 0.58 & 0.36 & 0.78 & $0.2\pm 0.1$ & $0.91^{+0.02}_{-0.02}$ \\
         & 65 & 3.1& 0.48 & 0.32 & 0.54 & 0.32 & 0.70& $0.1\pm 0.1$ & $0.91^{+0.03}_{-0.03}$\\
          & 75 & 3.47 & 0.42 & 0.31 & 0.47 & 0.29 &  0.63 & $0.1\pm 0.1$& $0.93^{+0.04}_{-0.04}$\\
    \hline
    \end{tabular*}
    {\raggedright \emph{Notes.} The third column gives the $y$-coordinate of the centre of the analysis region. The $x$ and $z$-coordinates are kept the same (i.e., $x=z=0$) for all the time instances and void sizes considered.
    \par}
    \label{tab:time_dependence}
\end{table*}

To check whether the turbulence properties depend on the time chosen at which it is analysed, we consider three different time instances at which there is a sufficiently large region of turbulence: $50\,\mathrm{ns}$, $65\,\mathrm{ns}$, and $75\,\mathrm{ns}$ after the shock initiation. The dimensions of the analysis region are kept constant for all the three times ($1.2\,\mathrm{mm}\,\times0.3\,\mathrm{mm} \,\times\,1.2\,\mathrm{mm}$). However, its position is moved upwards (along the positive $y$-direction) as time progresses, so as to select a region of well-developed turbulence at each time instance. Table~\ref{tab:time_dependence} lists the turbulent density dispersion, Mach number, and the driving parameter calculated at each time. We find that the overall shapes of the PDFs remain largely independent of time (not shown here). Although the spread of the density PDFs, as well as the Mach number PDFs, decreases slightly as time progresses, the driving parameter remains relatively unchanged. Thus, we conclude that the driving parameter in our case is largely independent of the choice of analysis time, as long as the analysis time and region are chosen such that the region contains largely fully-developed turbulent flow and is not directly affected by the shock (i.e., the shock has passed the analysis volume), and the analysis region is not affected by the boundaries of the setup.

\section{Numerical resolution study}
\label{sec:resolution}
Here we consider three different numerical resolutions for the standard case of $50\,\mu$m foam voids, and analyse the turbulence properties at $t=75\,\mathrm{ns}$ after the shock initiation. The dimension and position of the analysis region are kept constant for all the three cases. The resolutions considered are the standard resolution ($786 \times 1024 \times 786$ compute cells), half the standard resolution ($384\times 512\times 384$ compute cells), and double the standard resolution ($1536\times2048\times 1536$ compute cells). Table~\ref{tab:resolution} lists the quantities of interest. We see that the results are reasonably converged with the standard resolution that is used throughout this study.

While the turbulence driving parameter is relatively insensitive to the numerical resolution, we note that mixing of material in the post-shock medium will likely depend on the numerical resolution \citep[see e.g.,][]{BandaBarraganEtAl2020,BandaBarraganEtAl2021}. Noting that numerical viscosity starts acting on scales of $\lesssim 30$~grid cells \citep{KitsionasEtAl2009,SurEtAl2010,FederrathEtAl2011}, it is expected that mixing in the mid- and small-sized void cases is likely underestimated.

\begin{table*}
    \centering
    \caption{Numerical resolution study.}
    \def\arraystretch{1.4}
    \setlength{\tabcolsep}{0pt}
    \begin{tabular*}{\linewidth}{@{\extracolsep{\fill}}cccccccccccc}
    \hline
    Resolution & Number of compute cells & $\sigma_{\rho/ \langle \rho \rangle }$& $\sigma_{\mathcal{M}_x}$ & $\sigma_{\mathcal{M}_y}$ & $\sigma_{\mathcal{M}_z}$ & $\mathcal{M}$ & $\Gamma$ & $b$ \\
    \hline
        Low & $384\times 512\times 384$ & $0.26$ & $0.20$ & $0.29$ & $0.20$ & $0.41$ & $0.1\pm 0.1$ & $1.01\pm 0.07$\\
        Standard & $786 \times 1024 \times 786$ & $0.27$ & $0.22$ & $0.29$ & $0.22$ & $0.43$ & $0.1\pm 0.1$ & $0.99\pm 0.06$\\
        High & $1536\times2048\times 1536$ & $0.28$ & $0.21$ & $0.28$ & $0.21$ & $0.41$ & $0.1\pm 0.1$ & $1.07\pm0.07$ \\
    \hline
    \end{tabular*}
    \label{tab:resolution}
\end{table*}

\bsp	
\label{lastpage}

\end{document}